\documentclass[10pt]{iopart}
\usepackage{multicol}

\usepackage{natbib}


\usepackage{iopams}  
\usepackage{color}
\usepackage{tabularray}
\usepackage{graphicx}
\usepackage{epstopdf, epsfig}
\usepackage{lipsum}
\usepackage{multicol}
\usepackage{multirow}
\usepackage{graphicx}
\usepackage{textcomp}
\usepackage{rotating} 
\usepackage{verbatim}

\usepackage{array}

\usepackage{lineno}
\usepackage{xcolor}

\RequirePackage{colortbl}
\RequirePackage{booktabs}
\RequirePackage{algorithm}
\RequirePackage[noend]{algpseudocode}
\RequirePackage{changepage}

\usepackage{subcaption}
\usepackage{rotating}
\usepackage{adjustbox}
\usepackage{siunitx}
\UseTblrLibrary{booktabs}

\usepackage{pgfplots}
\pgfplotsset{compat=1.18}
\usepackage{tikz}

\newcommand{\ihat}{\textbf{\^\i}}

\begin{document}

\pagestyle{plain}

\title[Author guidelines for IOP Publishing journals in  \LaTeXe]{Tuna-Like Swimmers Experience a Fluid-Mediated Stable Side-by-Side Formation}

\author{Pedro C. Ormonde$^{1,3}$, Matthew Stasolla$^1$, Alec Menzer$^2$, Joseph Zhu$^2$, Hilary Bart-Smith$^2$, Haibo Dong$^2$, and Keith W. Moored$^1$}

\address{$^1$Department of Mechanical Engineering and Mechanics, Lehigh University, Packer Ave., Bethlehem,
PA 18015, USA}
\address{$^2$Department of Mechanical and Aerospace Engineering, University of Virginia, Charlottesville, VA 22904, USA}
\address{$^3$Center for Fluid Mechanics, School of Engineering, Brown University, Hope St., Providence, RI 02912,
USA}
\ead{pco@brown.edu}
\vspace{10pt}
\begin{indented}
\item[]April 2025
\end{indented}

\begin{abstract}
New free-swimming experiments and simulations are conducted on a pair of three-dimensional, bio-robotic swimmers composed of a body and tail section based on Yellowfin tuna, \textit{Thunnus albacares}. It is discovered that the pair converges spontaneously to a side-by-side schooling formation that is stable to perturbations in the swimming direction at a fixed lateral spacing. We reveal that for close lateral spacings of 43\% of the body length and thick, tuna-like bodies with a 22\% thickness-to-length ratio, the flow between the swimmers is accelerated in a ``channeling effect'' due to flow constriction. Consequently, this creates a low-pressure zone that is the primary mechanism generating a fluid-mediated restorative force, thereby making the side-by-side formation hydrodynamically stable. This quasi-steady mechanism makes the stability of the formation insensitive to the phase synchronization between the bio-robots in contrast to previous results for schooling foils. Moreover, in the side-by-side formation tuna-like swimmers are seen to have only a small reduction in their swimming speed and a concurrent small rise in their cost of transport. By leveraging this channeling effect, bio-robotic schools may be able to maintain a schooling formation with little or no control. This flow mechanism may also be present in biological schools of tuna-like fish where it may sculpt the formations observed in nature.
\end{abstract}

\vspace{2pc}

main.tex\maketitle
\ioptwocol

\section{Introduction}
Fish schooling is prevalent with an estimated 5,000 fish species schooling throughout their lives~\citep{Shaw1978} in schools ranging from just a few individuals to millions~\citep{partridge1982structure}. Schooling is known to affect a number of aspects of fish biology from social dynamics and reproduction~\citep{Shaw1978,kasumyan_schooling_2023,Neat2017} to protection from predation~\citep{breder1967,partridge1982structure,Godin1985} and improved foraging success~\citep{Pitcher1986,Ranta1990,wolf_schooling_1987}. It also affects the hydrodynamics of swimming, and it has been hypothesized that fish can save energy by swimming together in a school~\citep{weihs1973hydromechanics}. Until recently, the only data available in literature testing this hypothesis used indirect energetic estimates from observations such as tail beat frequency, heart rate, and aerobic energy~\citep{zhang2023energetics}. Now, definitive measurements directly quantifying total energy expenditure have verified that schooling can indeed provide substantial energy savings for schools swimming at moderate to high speeds~\citep{YangfanPLOS2024,ZhangLauder2024}. Moreover, the energy savings are thought to occur across a wide range of school sizes and species~\citep{parker1973reduced,thandiackal2023line}. 

Schools of real fish present numerous challenges to directly studying schooling energetics and their underlying hydrodynamic mechanisms. The highly dynamic nature of schooling formations~\citep{gerlotto_waves_2006,makris_critical_2009,handegard_method_2017} as well as large variability of swimming speeds, kinematics, size, and shape of  species~\citep{Burgerhout2013,abrahams_risk_1985,Hvas_2019} makes it difficult to extract a mechanistic understanding of the relevant hydrodynamic interactions occurring in schools. Physical bio-robotic and/or numerical fish-like swimmers can thus be studied as simplified models of real fish \citep{gravish2018robotics,lauder2022robotics}, one which researchers have control over the body shape/size, swimming kinematics, spatial formation, and temporal synchronization of the school. This allows for the salient flow physics to be retained and probed with independent variations of important hydrodynamic variables. 

Using numerical fish-like swimmers, performance benefits have been observed for schools~\citep{pan2022computational,kelly2023hydrodynamics}, where the hydrodynamic interactions between swimmers are affected by their body shape and school formation~\citep{KellyDong2024,Guo_phalanx2025}. However, these studies are performed for prescribed spatial arrangements where each swimmer has a fixed relative spacing and orientation. For a school of \textit{freely-swimming} fish-like swimmers to maintain a formation or cohesion at all, individuals must match their average swimming speed and direction over time. A cohesive stable formation may be achieved by continuous feedback control with kinematic adjustments by individuals or hydrodynamic interactions alone can produce restorative forces \textit{without any feedback control} leading to fluid-mediated stable formations, at least in the case of schooling hydrofoils~\citep{lighthill1975mathematical, newbolt2022lateral,baddoo2023generalization,han2023revealing}. In fact, one-dimensional, fluid-mediated streamwise stability was reported for in-line formations of schooling foils~\citep{ramananarivo2016flow,becker2015hydrodynamic,baddoo2023generalization} and two-dimensional fluid-mediated stability was found for foils in a side-by-side formation~\citep{ormonde2024two}. Oscillating hydrofoils act as idealized models of the caudal fin of a fish due to their hydrodynamic similarities~\citep{smits2019undulatory}, i.e. their comparable underlying added mass and circulatory forces as well as their similar wake structures characterized by the shedding of two alternating vortex rings per beating cycle~\citep{king2018experimental,buchholz2008wake}. 

The fluid-mediated stable formations found for schooling foils and the hydrodynamic similarities of foils to fish-like swimmers motivates us to investigate whether fluid-mediated stable formations \textit{extend to fish-like swimmers}. While many of the hydrodynamic interactions present in schools of hydrofoils~\citep{Lin_Wu_Zhang_Yang_2020,Lin_Wu_Zhang_Yang_2021,Gungor_Khalid_Hemmati_2022} may be relevant to fish-like bio-robots, no studies have examined the formation stability of \textit{three-dimensional bio-robotic swimmers} -- composed of a body and tail fin -- at Reynolds numbers comparable to that of real fish. The hydrodynamic interactions present in schools of bio-robots are known to be affected by body-body and wake-body interactions~\citep{Pan2024computational,kelly2023hydrodynamics,KellyDong2024,Guo_phalanx2025}, which are not present for schools of oscillatory foils~\citep{Boschitsch2014,kurt2018flow,Kurt2020,han2023revealing} due to the lack of a fore body. For this reason, the performance and stability of a school of fish-like bio-robots is largely unknown. If stable formations of bio-robots do exist, they provide an exciting flow physics pathway to maintain a cohesive formation with little or no control, which opens the door to potential energy savings for the swimmers.
 
In this study, we investigate the stability of canonical schooling formations for a pair of self-propelled, fish-like bio-robots through water channel experiments and numerical simulations. The two swimmers are allowed to move freely in the stream-wise direction and are constrained in the cross-stream direction so the dynamic behavior of the schooling pair can be studied. The synchronization between the caudal fins of the bio-robots is set to both in-phase and out-of-phase motions. Additionally, we study the effects of body thickness on the hydrodynamics of the school. Flow structures, hydrodynamic performance, and streamwise stability are all examined for the school. In Section~\ref{sec:methods} we describe the experimental methods and numerical simulations. Section~\ref{sec:results} presents results on the stability of staggered and side-by-side formations, describing the hydrodynamic mechanism that leads to a stable side-by-side formation for a tuna-like bio-robot. The hydrodynamic performance of the school as a function of the streamwise spacing between bio-robots is also presented. Section~\ref{sec:conclusions} discusses the results and their implications to schooling fish-like bio-robots.

\section{Methods} \label{sec:methods}

\begin{figure}
    \centering
    \includegraphics[width=\linewidth]{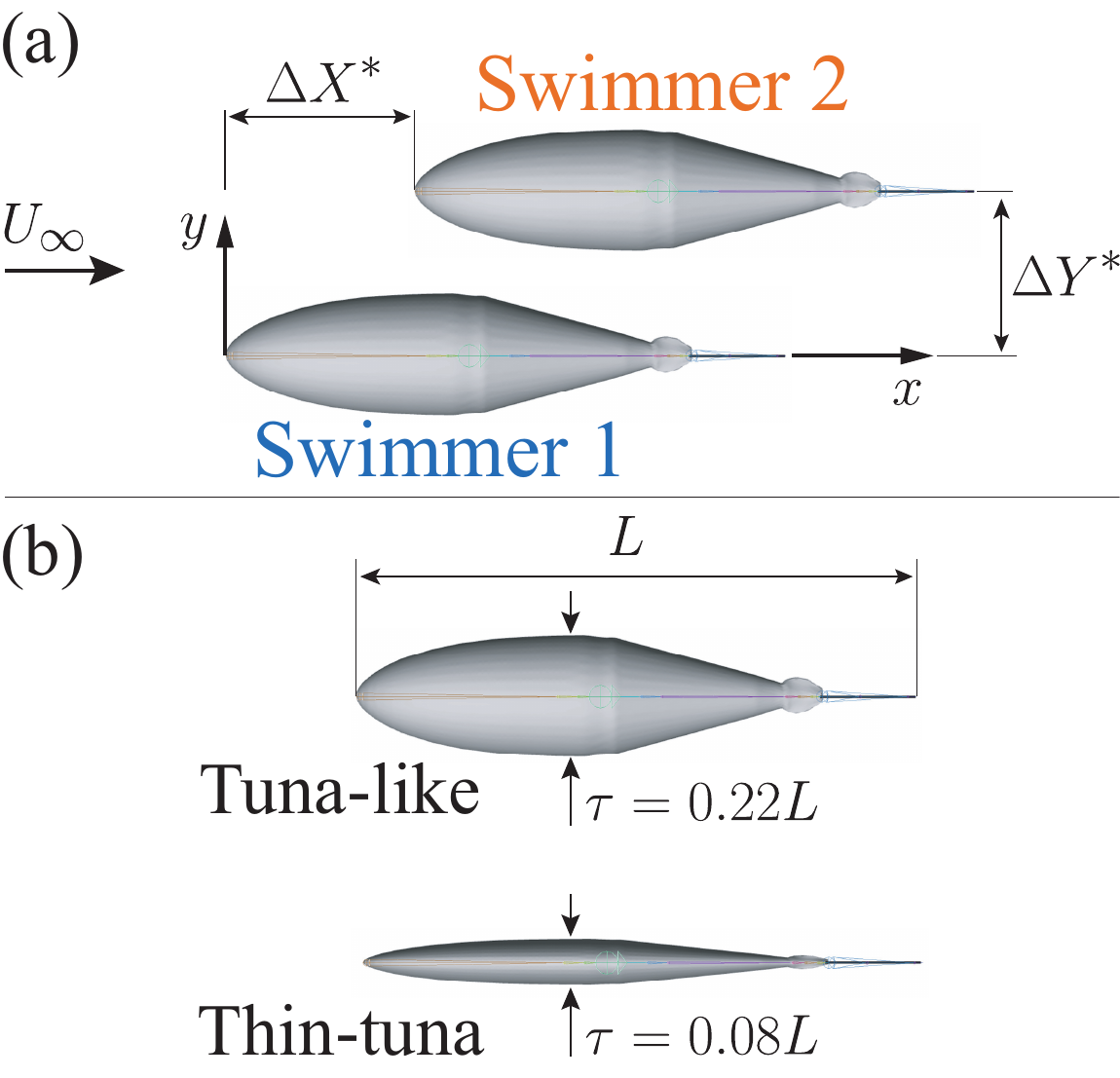}
    \caption{(a) Definition of coordinate system and non-dimensional distance between swimmers in the streamwise, $\Delta X^*=(x_2-x_1)/L$, and cross-stream, $\Delta Y^*=(y_2 - y_1)/L$, directions. (b) Top view of tuna-like and thin-tuna geometries and their respective maximum thickness to body length ratio $\tau/L$.}
    \label{fig:coords_thickness}
\end{figure}

Water channel experiments and numerical simulations were performed for pairs of fish-like bio-robots. The $x-y$ coordinate system is nondimensionalized by the total length of the bio-robot, $L$, defining $x^* = x/L$ and $y^* = y/L$ as the non-dimensional streamwise and cross-stream directions, respectively. The adopted coordinate system is described in Figure~\ref{fig:coords_thickness}(a) along with the definition of the streamwise position of swimmer 2, $x_2$, relative to the position of swimmer 1, $x_1$, as $\Delta X^*= (x_2 - x_1)/L$. The relative cross-stream distance between them is $\Delta Y^* = (y_2-y_1)/L=0.43$, measured from the body centerline. Two different body geometries were used in the study. The baseline case is inspired on the morphological features of the yellowfin tuna, \textit{Thunnus albacares}. It is named as the \emph{tuna-like} swimmer, or \emph{Tuna1} and \emph{Tuna2} throughout this study and has a maximum thickness $\tau/L = 0.22$ in the cross-stream direction. A second swimmer shape is also studied, named as \emph{thin-tuna}, or \emph{Thin1} and \emph{Thin2}. Its baseline body geometry is modified by scaling down the dimensions of the tuna-like swimmer in the cross-stream direction $y$. The thin-tuna body has a maximum non-dimensional thickness of $\tau/L=0.08$.

\subsection{Experimental setup}
\begin{figure*}
     \centering
         \includegraphics[width=\textwidth]{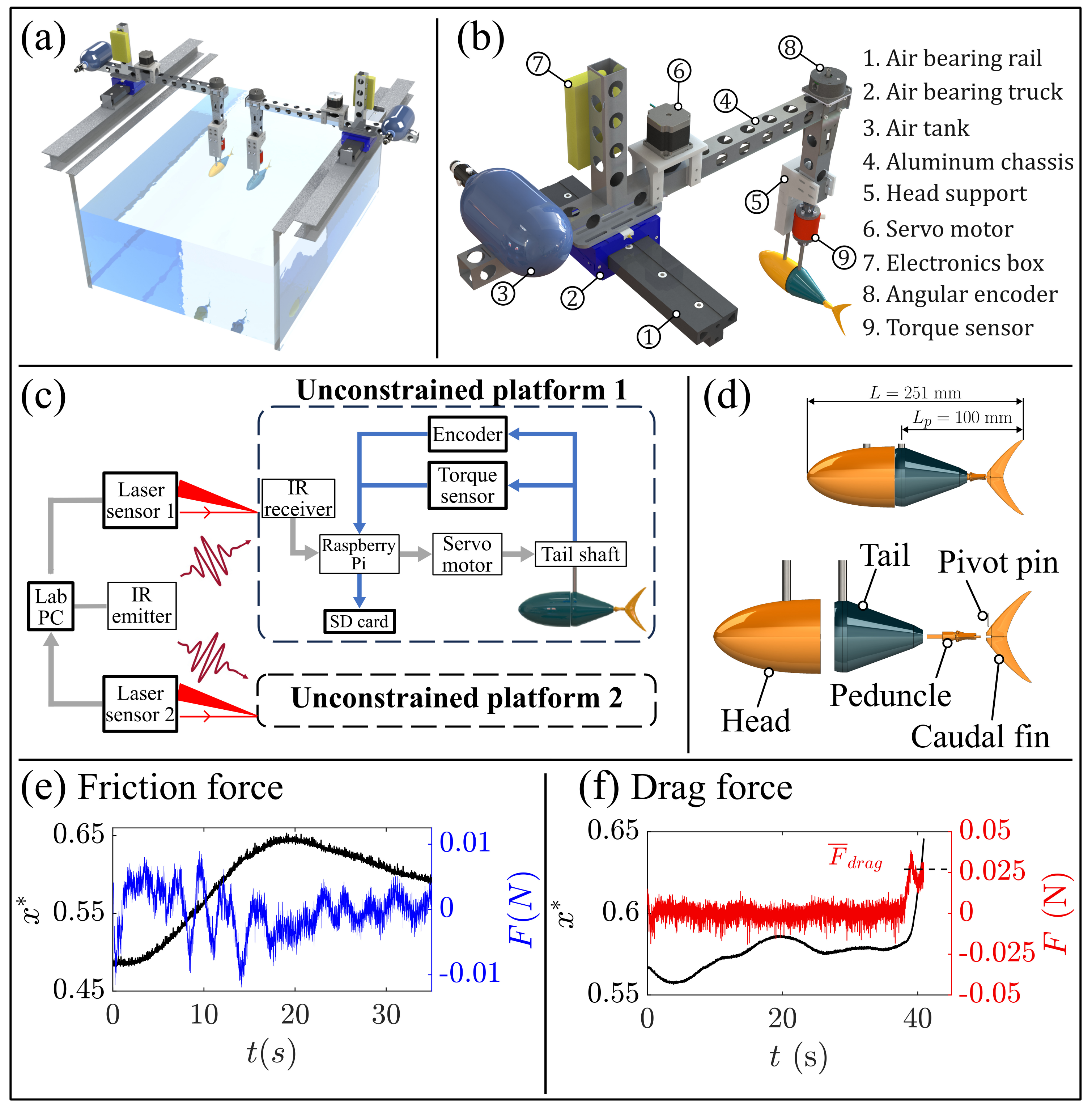}
         \caption{(a) Isometric view of the experimental setup. (b) Schematics of the air-bearing rail and robotic platform. (c) Schematics of the wireless actuation and embedded data acquisition. Laser distance sensors measure the position of each swimmer in time. Infrared (IR) pulses control the kinematics remotely. Unconstrained platforms 1 and 2 are identical. Grey arrows represent flow of actuation signals, and the blue arrows represent flow of data measurement signals. The torque data is stored locally by an SD card and retrieved over WiFi at the end of the experiment. The position of the bio-robots and measured torque signals can be synchronized in post-processing using the IR pulses as a reference clock. (d) Detailed view of the bio-robot tail, peduncle and caudal fin. (e) Air-bearing spurious force (blue line) calculated from the trajectory of the carriage (black line) at zero flow. (f) Drag force $\overline{F}_{\text{drag}}$ calculated from the trajectory of the carriage (dashed black line) for the bio-robot at rest subject to the freestream $U$.}
    \label{fig:exp_setup}
\end{figure*}

Experiments were performed in a recirculating water channel with test section dimensions of 2.43 m in length $\times$ 0.91 m in width $\times$ 0.61 m in depth. Figure \ref{fig:exp_setup}(a) shows the experimental setup for two fish-inspired bio-robots submerged in the water. The bio-robots are each connected to their respective robotic platform that is supported by an air-bearing rail mounted on the side walls of the water channel. The two platforms are identical. The rails are aligned with the free-stream flow direction ($x$-axis) and air-bearing trucks run along it with very low friction, allowing the bio-robots to move freely in the streamwise direction. The cross-stream direction is constrained. 

Figure~\ref{fig:exp_setup}(b) shows a detailed schematic of the robotic platform supported by the air-bearing truck. An aluminum chassis is rigidly secured to the aluminum truck and supports all of the remaining components: an air tank, an electronics box, a servo motor, two sensors, and the bio-robot body and tail. The actuation system and its batteries are embedded on the unconstrained platform to prevent any wires from introducing external forces to the system, which could corrupt the results from the unconstrained, self-propelled swimmers. 

The fish-like bio-robots are constituted of two main parts: A drag-producing ``head", and a ``propulsive tail section''. The head is rigidly attached to the aluminum chassis and does not move relative to the carriage/truck. The propulsive tail section is driven by the rotating shaft in an oscillatory motion that generates a propulsive force for the bio-robot. Figure \ref{fig:exp_setup}(d) shows the head and the components of the propulsive tail section, located on the back-half of the body. The propulsive section is comprised of a tail, keel, caudal fin, and metal springs. The caudal fin is attached to the peduncle through a pivot pin that allows for passive rotation. The metal springs are made of stainless steel wires and determine the stiffness of the joint. The length of the propulsive tail section is $L_p=100$ mm, measured from the shaft axis to the tip of the caudal fin, and the caudal fin has a span $b=87$\ mm, planform area $S=1.1\times10^{-3}\,\text{m}^2$ and aspect ratio $AR = b^2/S = 6.9$. The total bio-robot length is $L=251$ mm. The sinusoidal motion of the servo motor is transmitted to the stainless steel tail shaft by a belt. The prescribed sinusoidal motion of the tail is $\theta_1(t) = \theta_0 \sin{\left(2\pi f t\right)}$ for swimmer 1, and $\theta_2(t) = \theta_0 \sin{\left(2\pi f t + \phi\right)}$  for swimmer 2, where $\theta_0$ is the pitching amplitude of the tail in degrees, $f$ is the oscillation frequency, and $\phi$ is the phase synchronization between swimmers 2 and 1. An in-phase synchronization is $\phi=0$ and an out-of-phase synchronization is $\phi = \pi$. The non-dimensional peak-to-peak amplitude of the caudal fin is defined as $A^* = A/L = 2\left(L_p/L\right) \sin\theta_0 $, where $A$ is the peak-to-peak tail amplitude. The Reynolds number based on the total length is defined as $Re = U L/\nu$, where $U$ is the swimming speed and $\nu$ is the kinematic viscosity of water.

Figure \ref{fig:exp_setup}(c) presents the schematics of the actuation mechanism and real-time data acquisition. Wireless synchronization of the two independent motors (one for each bio-robot) is achieved via infrared (IR) communication. A laboratory computer controls an infrared emitter that generates communication pulses. On the unconstrained platform, the command pulses are captured by an infrared receiver that sends the pulses to the micro-controller. These pulses are used as start-stop commands and as a clock signal. The clock signal synchronizes the motion of the two bio-robot tails and maintains a constant phase $\phi$ in time. The laboratory computer reads data from laser distance sensors that measure the position of the bio-robots. The measured signal is referenced in time using the IR timing pulses. The tail shaft of the bio-robot is instrumented with an absolute analog encoder that measures its angular position (US Digital A2K optical encoder) with 1$^{\circ}$ rms of output noise. A reaction torque sensor (FUTEK FSH03984 coupled to a strain gage analog amplifier FSH03864) is directly attached to the pitching shaft. The shaft position and reactive torque are measured in real-time during unconstrained, self-propelled experiments. The analog output signals from both instruments are sent to an on-board data acquisition card (MCC118 DAQ) that is directly attached to the micro-controller. The data is saved to the micro-controller SD card and is retrieved over a local area network (LAN) by accessing the micro-controller over a secure shell protocol (SSH) at the end of the experiment. The stream-wise position data is synchronized to the torque data based on the IR clock signal. The total instantaneous power input $P(t) = M_{\theta}\dot{\theta}$ is calculated from the pitching moment $M_{\theta}$ and angular velocity $\dot{\theta}$ of the tail shaft. The power coefficient $C_P$ is then calculated from $P(t)$ as 

\begin{equation} \label{eq:Cp}
C_P = \frac{\overline P}{\frac{1}{2}\rho {U}^3 S}
\end{equation}

\noindent where $\overline{P}$ is the time-averaged power and $\rho$ is the fluid density. The cost of transport is defined as

\begin{equation} \label{eq:CoT}
	CoT = \frac{\overline{P}}{m_{\text{exp}} U}
\end{equation}
\noindent where $m_{\text{exp}}=6.8$ kg is the total mass of the bio-robotic swimmer.\\
	
\subsubsection{Air-bearing rail leveling and friction}
\label{subsubsec:rail}
The air-bearing system introduces two types of extraneous forces to the unconstrained bio-robot: friction and forces caused by rail levelness and manufacturing tolerances. A machinist level with $0.0024^\circ$ accuracy is used to level the rails. An additional procedure is performed to quantify the external forces generated by the air-bearing system, where the truck is set to rest at several locations along the rails while air is supplied to the bearings. Due to levelness and manufacturing imperfections the truck moves by itself. The motion of the truck is then tracked and used to calculate its acceleration. This acceleration, multiplied by the total suspended mass of the system ($m = 6.3$ kg for these tests) is the total spurious force introduced by the air-bearing. The maximum calculated force is $0.01$ N. Figure~\ref{fig:exp_setup}(e) shows the time-varying trajectory of the air-bearing carriage (black line) and the estimated spurious forces (blue line). This force is then compared to the drag force experienced by the fish-like bio-robot at rest (no tail motion) under the lowest speed tested. Figure~\ref{fig:exp_setup}(f) shows the trajectory of the bio-robot during an experimental trial (black line). At the end of the trial ($t_{end}$ in the horizontal axis) the bio-robot stops and the drag force moves it downstream ($+x$ direction). The drag force is then calculated to be $\overline{F}_{\text{drag}} = 0.03$ N. While not negligible, the maximum spurious forces generated by the air-bearing rail are found to be at least three times lower than the minimum drag force experienced by the bio-robot for the lowest freestream speed. 

\subsubsection{Characterization of bio-robot hydrodynamics}

\begin{figure}[b]
	\centering
	\includegraphics[width=\linewidth]{"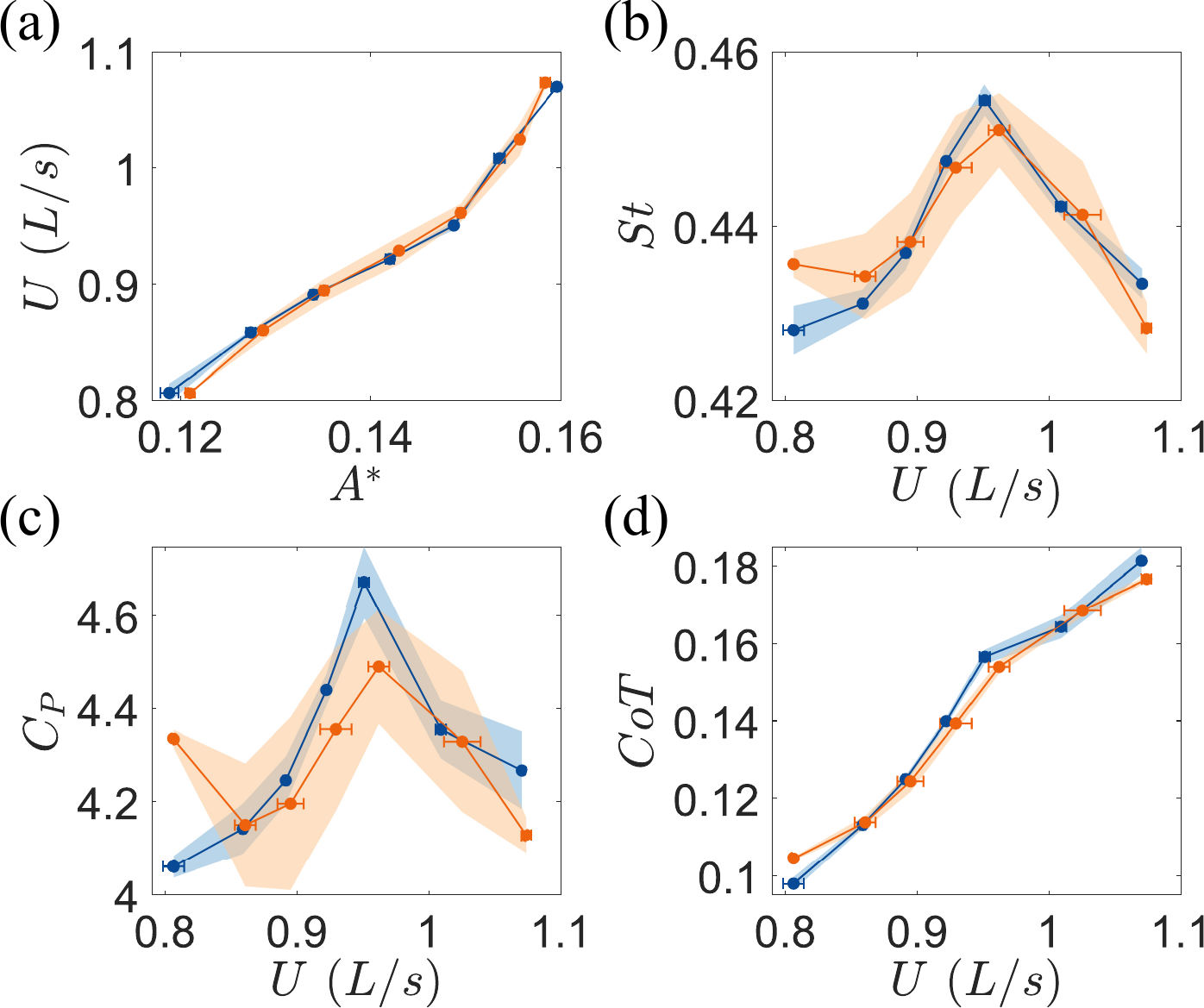"}
	\caption{Hydrodynamic performance of the tuna-like bio-robots swimming in isolation. Blue: Tuna1; Orange: Tuna2. All data is averaged over three trials. Shaded area indicates the standard deviation of the vertical axis; horizontal bars represent the standard deviation of the horizontal axis. (a) Time-averaged swimming speed (cruising speed) in body lengths per second as a function of the dimensionless tail beat amplitude. (b) Strouhal number  as a function of swimming speed. (c) Power coefficient as a function of swimming speed. (d) Cost of transport as a function of swimming speed.}
	\label{fig:iso_performance}
\end{figure}

The swimming speed, power, and cost of transport of the bio-robots swimming in isolation are determined for a range of dimensionless amplitude and a fixed tail beat frequency, $f = 2.9$ Hz. In free, non-accelerating swimming, the time-average thrust and drag forces are equal in magnitude and define the time-average swimming speed, also known as the cruising speed. Video recordings of the bio-robots were used to measure the phase lag between the heaving and pitching kinematics of the caudal fin, measured to be $82 \pm 2^{\circ}$. This phase lag is due to the springs connecting the caudal fin to the peduncle. The caudal fin pivots freely around the pivot pin, as depicted in Figure~\ref{fig:exp_setup}(d). Figure~\ref{fig:iso_performance}(a) shows the output swimming speed $U$ in body-lengths per second ($L$/s) as a function of the  amplitude for the two tuna-like bio-robots, each swimming in isolation. The Reynolds number range achieved on these experiments is $Re = 5.3 - 7.0 \times 10^4$. Figure~\ref{fig:iso_performance}(b) shows the Strouhal number $St = f A/U$ as a function of the swimming speed. The measured Strouhal number is relatively insensitive to the prescribed amplitudes and swimming speeds obtained. For the prescribed amplitudes ranging from 0.12 to 0.16, a change of 33\%, the maximum measured variation in the Strouhal number was 7\%. This is in agreement with a relatively linear response of the swimming speed as a function of $A^*$ observed for pitching hydrofoils~\citep{moored2019inviscid}. The time-averaged power coefficient and cost of transport are shown in Figures \ref{fig:iso_performance}(c) and \ref{fig:iso_performance}(d), respectively. The power coefficient peaks at $U = 0.95 \ L$/s, whereas the cost of transport grows monotonically with increasing swimming speed.

\subsubsection{Experiments of two interacting bio-robots}
Experiments with two bio-robots swimming simultaneously were performed at a fixed cross-stream separation distance of $Y^*=0.43$, the smallest allowable distance for the setup. The relative position between the two bio-robots was measured by the laser distance sensors. The prescribed tail beat frequency was set to $f=2.9$ Hz for all cases, and two phase synchronizations $\phi=0$ and $\pi$ were used throughout the experiments. As shown in Figure~\ref{fig:iso_performance}(a), there are small differences in the output swimming speed of the two bio-robots for a given tail amplitude. For that reason, in all schooling experiments their tail beat amplitudes are tuned such that both bio-robots achieved the same isolated swimming speed $U = 0.82$ ($L$/s). Tuna1 is prescribed an amplitude $A^*_1=0.11$ while Tuna2 is prescribed a slightly higher amplitude $A^*_2=0.12$. 

\subsection{Computational modeling and numerical simulations}

\begin{figure}[tp]
	\centering
	\includegraphics[width=1\linewidth]{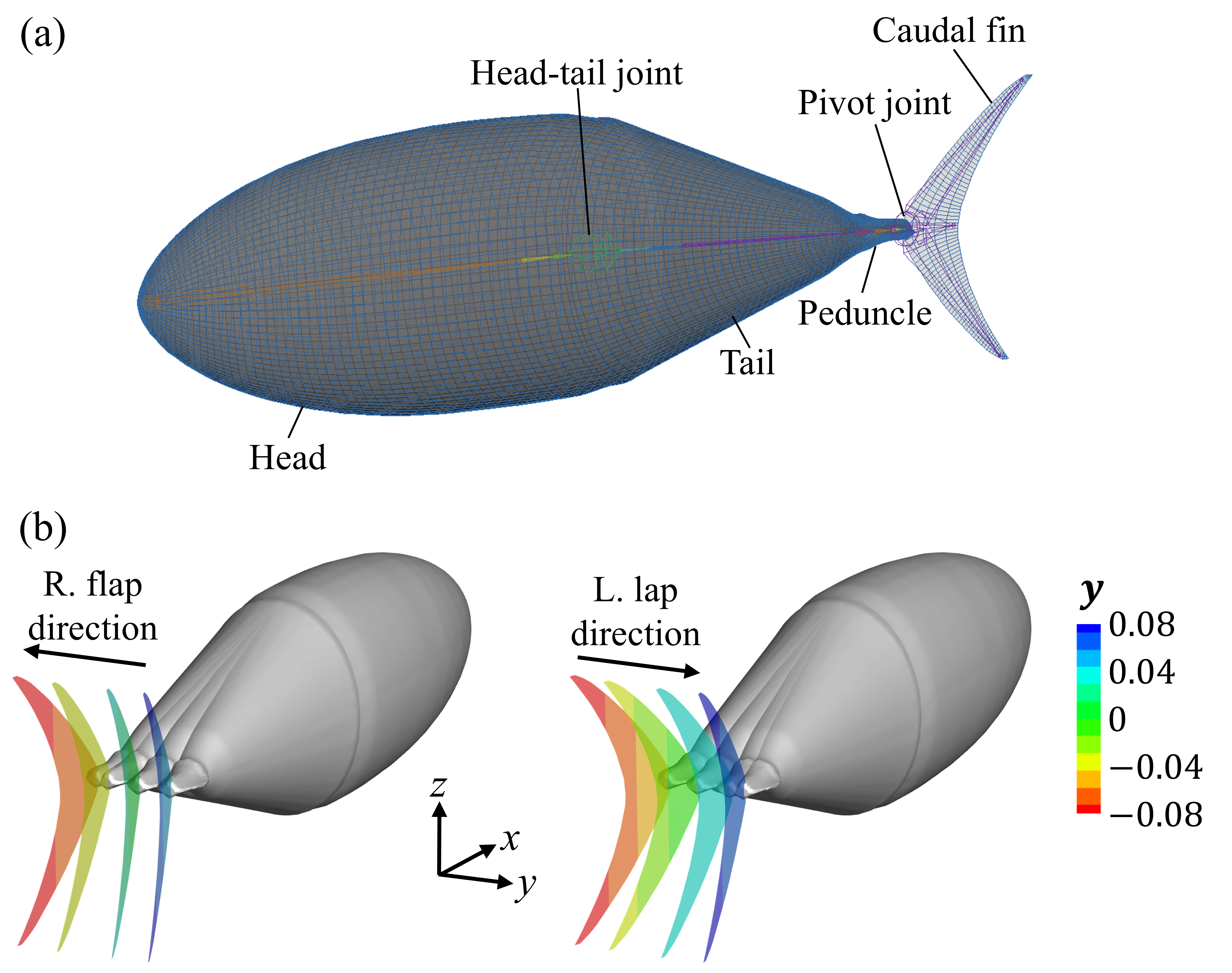}
	\caption{(a): Illustration of the surface meshes and virtual skeleton used to assign the kinematics to the computational model. The ``head-tail'' and ``pivot'' joints align with the locations at which the tail and caudal fin surfaces rotate on the tunabot model in Figure ~\ref{fig:exp_setup}(d). (b) Close-up view of the prescribed kinematics during the right (R) flap and left (L) flap that align with the experiments. The ``head-tail'' joint angle was prescribed as an 18$^{\circ}$ amplitude (peak-to-peak) sinusoidal function. The ``pivot'' joint angle was assigned a 30$^{\circ}$ amplitude (peak-to-peak) sinusoidal function that lagged the ``head-tail' joint by 90$^{\circ}$ thereby producing the tuna-like flapping motion.}
	\label{fig:comp_model}
\end{figure}

The body and tail morphology in the computational model is derived directly from the files used to manufacture the bio-robot body, tail, peduncle, and caudal fin surfaces used in the experiments. In the computational model in Figure~\ref{fig:comp_model}(a), the tail, head, and peduncle are all modeled using one continuous solid surface mesh, while the caudal fin is modeled by a separate membrane surface mesh. Using Autodesk Maya\textregistered, a virtual skeleton is assigned such that the locations of actuation on the bio-robots used on the experiments (Figure~\ref{fig:exp_setup}(d)) align with the virtual joint locations. Scaling is conveniently applied to the surface mesh of the baseline, tuna-like body and tail to produce the ``thin-tuna" body geometry, as described in Section~\ref{sec:methods}. Figure~\ref{fig:comp_model}(a) illustrates the surface meshes and virtual skeleton in the virtual environment. Rotations are applied to the ``head-tail'' joint (akin to rotating the tail, see Figure~\ref{fig:exp_setup}) and to the ``pivot'' joint between the peduncle and caudal fin so that the flapping motion of the virtual model can closely align to that observed in the experiments. The tail and caudal fin kinematics are visualized over a cycle of motion in Figure~\ref{fig:comp_model}(b), while the head motion is fixed (as in the experiments).

\begin{figure}[tp]
	\centering
	\includegraphics[width=1\linewidth]{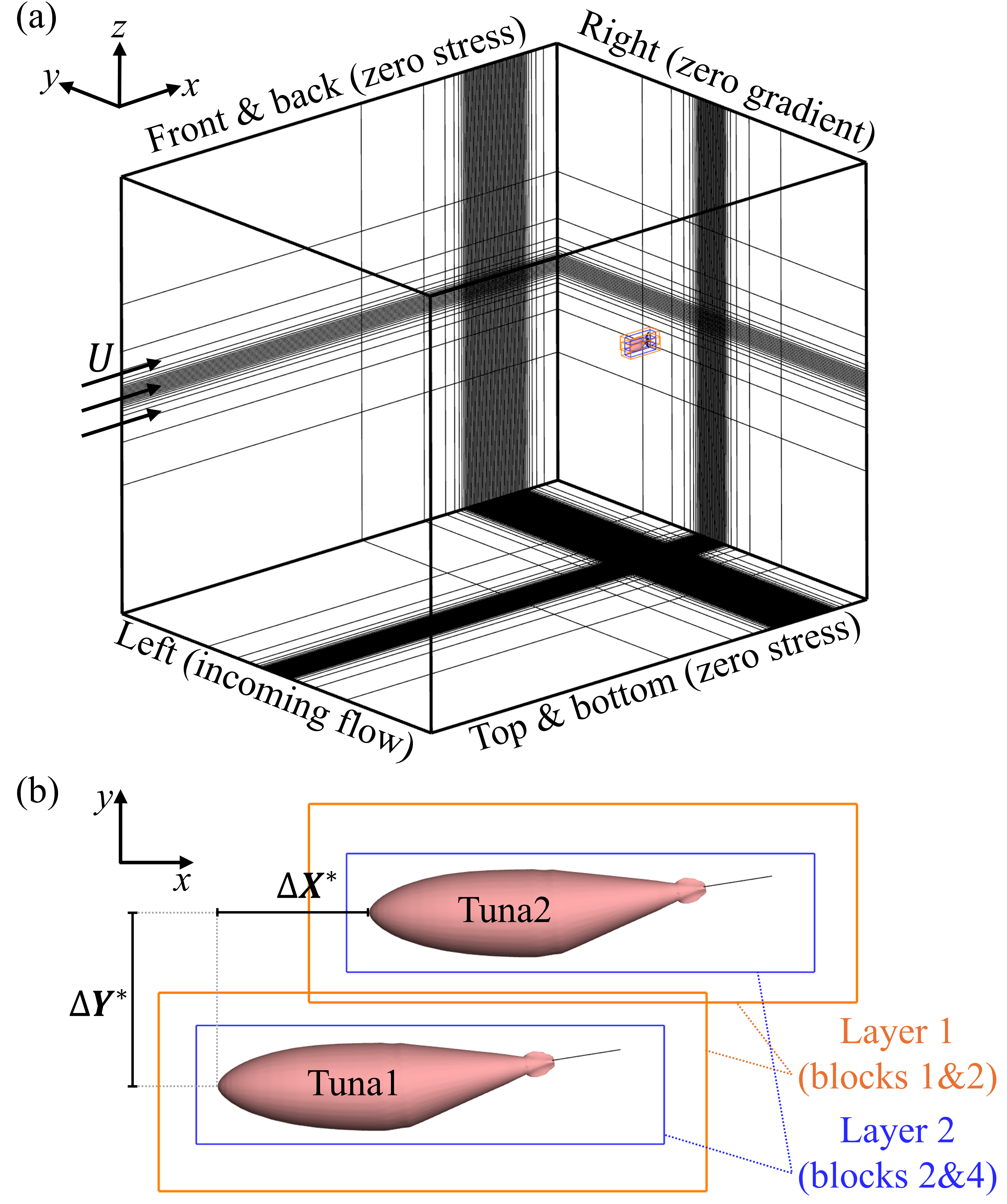}
	\caption{Schematic of the computational domain. (a) Illustration of grids (every 4th is shown to improve visibility) and velocity boundary conditions on the domain boundaries. (b) Close-up view of the mesh refinement regions surrounding each body, along with the definition of the spatial arrangement in the simulations.}
	\label{fig:comp_domain}
\end{figure}

The numerical simulations allow us control over the non-dimensional mass of the bio-robot $m^* = m/m_{f}$, where $m$ is the bio-robot's mass and $m_{f}$ is the mass of fluid displaced by the total volume of the bio-robot. Due to practical limitations, the total mass of the bio-robots in the experimental setup is relatively large compared to the typical value of $m^*\approx1$ for real fish, which affects the dynamic behavior of the system. In the simulations, on the other hand, we are able to achieve $m^*=1$, which is more representative of fish biology. Another use of the simulations is to facilitate the visualization of the three-dimensional flow structures and quantification of pressure fields surrounding the swimmers, from which we can determine the mechanistic sources of the hydrodynamic interactions between the pair, and extract performance metrics not available from the experiments. The governing equations in the computational fluid dynamics (CFD) simulations are the incompressible Navier-Stokes equations, shown in Equation~\ref{eq:navstokes}, where $u_{i}$ are the velocity components ($i = 1,2,3$ for $x$, $y$, and $z$ respectively), $t$ is time, $p$ is pressure, and $\rho$ is the fluid density. All flow variables are expressed in the appropriate non-dimensional forms. The selected Reynolds number for the CFD simulations is $Re=5407$, an order of magnitude smaller than that of the experiments. This is adopted to reduce computational cost, which would be prohibitively large for $O(Re) = 10^4$ to match the Reynolds number of the experiments. Previous studies have established that flows at $O(Re) = 10^3$ replicate the key features of wake structures and flow interactions of fish-like swimming at Reynolds numbers in the range of $O(Re) = 10^4 -- 10^5$~\citep{Buchholz2006,Menzer2025}. A schematic depicting the computational domain is shown in Figure~\ref{fig:comp_domain}.

\begin{equation}
    \label{eq:navstokes}
        \frac{\partial u_i}{\partial x_i} = 0; \hspace{0.1 in} \frac{\partial u_i}{\partial t} + \frac{\partial u_i u_j}{\partial x_j} = -\frac{\partial p}{\partial x_i} + \frac{1}{Re}\frac{\partial^2 u_i}{\partial x_j \partial x_j}
\end{equation}

The Navier-Stokes equations are discretized using a cell-centered collocated arrangement of the flow variables on a non-conformal Cartesian background mesh. A fractional step method is used for time stepping of equation~\ref{eq:navstokes}. Diffusive terms are discretized by an implicit Crank-Nicolson scheme and convective terms are discretized by an Adams-Bashforth scheme. The velocity boundary conditions ensure incoming flow on the left, zero stress on the top, bottom, front, and back boundaries and zero gradient outflow on the right (see Figure~\ref{fig:comp_domain}(a)). Pressure boundary conditions on all boundaries is the homogeneous Neumann boundary condition. 

A sharp-interfaced immersed-boundary method accurately enforces the boundary conditions on the fish-like bio-robot surfaces \citep{Mittal2008}. Fine grids are therefore needed to accurately resolve the near-body flow fields and forces on the body. To this end, the flow solver utilizes a tree-topological local mesh refinement technique. Regions of mesh refinement (referred to as blocks) are nested within each other to double the grid density in each direction (see Figure~\ref{fig:comp_domain}(b))~\citep{WZhang2023}. Layer 1 refines the base layer (block 0). Layer 1 (blocks 1,2) contains two connected larger blocks, ensuring the flow patterns between the two robotic fish is accurately simulated. Layer 2 (block 3,4) further reduces the grid immediately surrounding the robotic fish to ensure that the near-body vortex structures and force calculations are accurate. The governing equations are solved in parallel on each of these blocks, allowing for an efficient solution of the flow field and forces. This refinement technique yields a uniform grid density of $\Delta = 0.005 L$ in each direction. A similar grid resolution has previously been shown to be sufficient for studying flows around fish-like bodies~\citep{Liu2017}. Free swimming capabilities of the solver are also utilized, where the imbalanced forces experienced by swimmer 1 and swimmer 2 result in accelerations of the bodies. 
The block boundaries track the motion of the fish to ensure that refinement blocks are able to contain the moving bodies. When swimmer 1 or swimmer 2 are detected near the boundary of the block that encloses it, the block is incrementally moved away from the surface in the next time step, ensuring proper interpolation of flow information ahead of and behind the block~\citep{WZhang2023}. The cross-stream distance between swimmers and free-stream velocity in the simulations (Figures~\ref{fig:comp_domain}(a) and~\ref{fig:comp_domain}(b)) are the same as in the water channel experiments.

The net thrust coefficient of each swimmer is defined as $C_T = \overline{T}/(\frac{1}{2}\rho U^2 S)$, where $\overline{T}$ is the time-averaged net force in the streamwise direction and $S$ is the planform area of the caudal fin. The normalized thrust coefficient is defined as $C^*_{T} = C_T/C_{T,\text{iso}}$, where $C_{T,\text{iso}}$ is measured for a swimmer in isolation.

\section{Results}
\label{sec:results} 
The stability of a schooling formation is investigated under the hypothesis that self-induced hydrodynamic interactions change the performance of each swimmer, which in turn may create restorative streamwise forces that repel/attract the two swimmers from/to one another. Hydrodynamic interactions between two oscillatory hydrofoils with an out-of-phase synchronization have been shown to spontaneously form a side-by-side schooling formation~\citep{ormonde2024two} that can sustain itself even in the presence of external disturbances. Inspired by this finding, we explore the hypothesis that restorative hydrodynamic interactions between a pair of fish-like bio-robotic swimmers may also exist, particularly by altering the propulsive performance of the bio-robots.  

\subsection{Spontaneous side-by-side formation of schooling bio-robots} \label{subsec:self-organization}

\begin{figure*}[t!]
	\centering
	\includegraphics[width=1\linewidth]{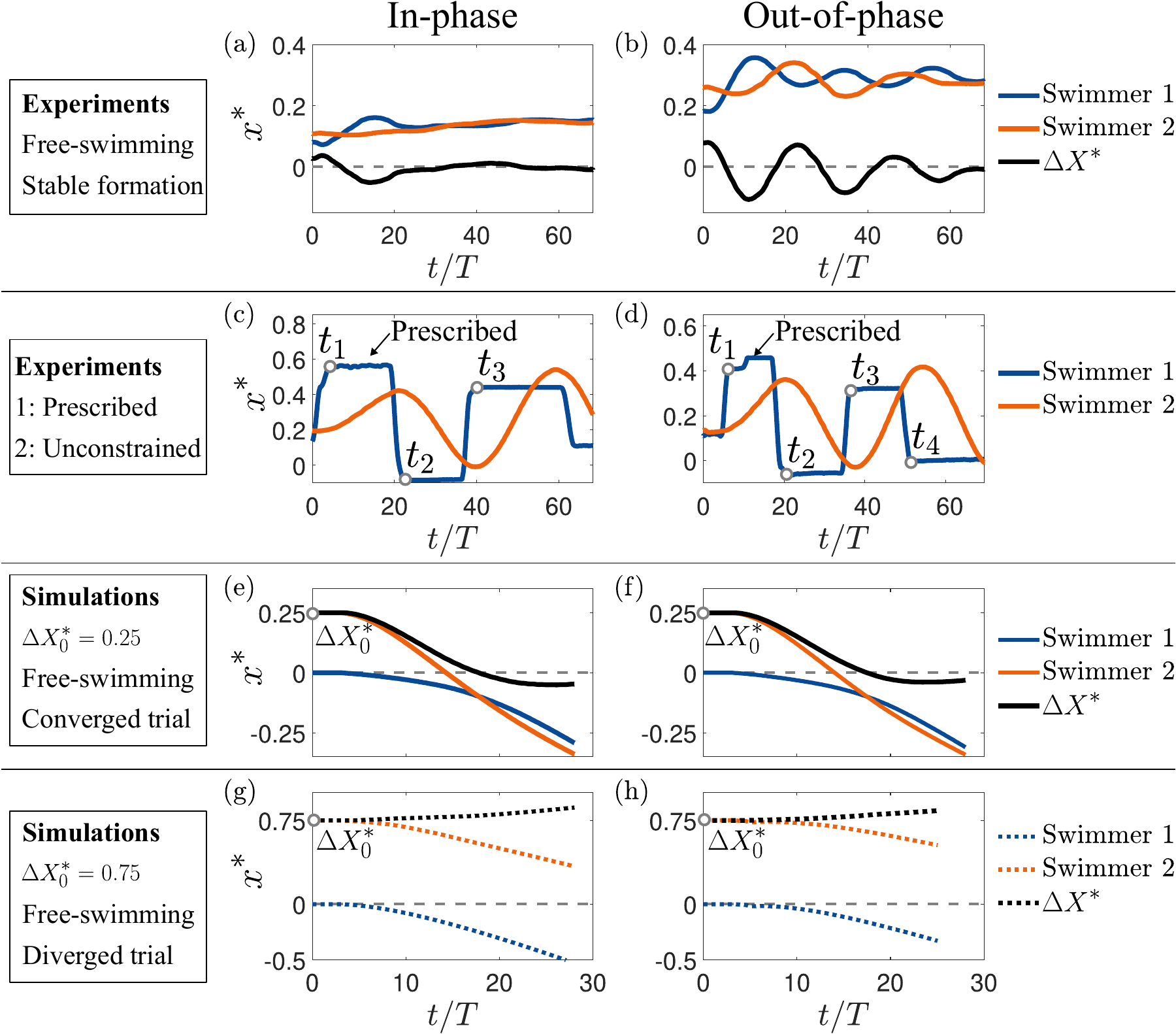}
	\caption{Trajectories of two tuna-like swimmers for a fixed cross-stream separation distance $\Delta Y^* = 0.43$. Blue line: swimmer 1; Orange line: swimmer 2; Black line: $\Delta X^*(t)$. (a,b) Experimental trajectories of freely swimming (unconstrained) swimmers reaching stable side-by-side formations, for an in-phase and out-of-phase synchrony, respectively. (c,d) Experimental, free-swimming trajectories of swimmer 2 in response to manually prescribed (perturbed) position of swimmer 1, for an in-phase and out-of-phase synchrony, respectively. (e,f) Trajectories of swimmers 1 and 2 from unconstrained simulations. Initial condition $\Delta X^*_0=0.25$ leads to a converged side-by-side formation for both synchronies. (g,h) Trajectories of Swimmers 1 and 2 from unconstrained simulations. Initial condition, $\Delta X^*_0=0.75$, increases the distance between the pair, leading to a diverging case. A side-by-side formation is not achieved.}
	\label{fig:trajectories}
\end{figure*}

Two tuna-like bio-robots were positioned close to each other in the water channel and allowed to swim freely and independently in the streamwise direction. Figure~\ref{fig:trajectories}(a) shows the trajectories of Tuna1 (blue line) and Tuna2 (orange line) starting with Tuna1 0.03 $L$ upstream of Tuna2, initially positioned at an arbitrary location along the rails with an in-phase synchronization. Observe how their separation distance represented by the black line converges to $\Delta X^*_{eq}=0$, i.e., a side-by-side schooling formation, despite the change in their absolute position over time. A second experiment, depicted in Figure~\ref{fig:trajectories}(b), shows Tuna1 initially positioned 0.08 $L$ upstream of Tuna2, a separation distance 2.5 times larger than that of Figure~\ref{fig:trajectories}(a), and with an out-of-phase synchronization. The two bio-robots move upstream and downstream, and their relative position (black line) shows a damped oscillatory behavior where the individuals alternate between being upstream and downstream of their neighbor. After 60 cycles, the system converges to a side-by-side formation $\Delta X^*_{eq}=0$. This is the same behavior observed in Figure~\ref{fig:trajectories}(a), however with a larger settling time. These results show the spontaneous self-organization of the pair of tuna-like swimmers into a side-by-side schooling formation. The side-by-side formation is thus a stable equilibrium state of the system of two bio-robots, and the number of cycles required to achieve it depends on the initial streamwise separation distance between swimmers. Larger streamwise distances lead to a higher number of cycles required for convergence, and \textit{vice-versa}.

Further experiments were conducted to confirm that the stable equilibrium formation observed on Figures~\ref{fig:trajectories}(a) and~\ref{fig:trajectories}(b) results from
hydrodynamic interactions between the two bio-robots, and are not caused by spurious forces due to the experimental setup (as discussed in Section~\ref{subsubsec:rail}). For these experiments, depicted on Figures~\ref{fig:trajectories}(c) and~\ref{fig:trajectories}(d), the position of Tuna1 is prescribed, i.e. Tuna1 is not allowed to swim freely, while Tuna2 is unconstrained in the streamwise direction and free to swim. In Figure~\ref{fig:trajectories}(c) the two swimmers start nearly side-by-side at $t/T = 0$ at an arbitrary streamwise location along the two rails with an in-phase synchronization. At instant $t_1$, Tuna1 is manually moved downstream and held at $x^* = 0.56$, and Tuna2 is observed to move downstream towards Tuna1, in response to its new relative position. Next, at instant $t_2$, Tuna1 is manually moved upstream to a new position $x^* = -0.08$, upstream of Tuna2. Once more, Tuna2 spontaneously moves towards Tuna1, swimming upstream. The same behavior is observed for an out-of-phase synchrony, as depicted in Figure~\ref{fig:trajectories}(d): Tuna2, the freely-swimming bio-robot, tracks the position of its neighbor \textit{with no control}. Movie 1, found in the supplementary material, shows an example of Tuna2 spontaneously moving according to the manually prescribed position of Tuna1.

Simulations of unconstrained tuna-like swimmers were also performed for the same lateral separation distance. Figures~\ref{fig:trajectories}(e) and~\ref{fig:trajectories}(f) show the resulting trajectories for an in-phase and out-of-phase synchronization, respectively, with an initial separation distance $\Delta X^*_0 =0.25$. Similar to the behavior observed in the experiments, the schooling formation converges to a stable side-by-side formation $\Delta X^*\approx0$ (shown by the black solid lines) after roughly 25 tail beat cycles. Unconstrained simulations were also performed for a larger initial separation distance $\Delta X^*_0= 0.75$, shown in Figures~\ref{fig:trajectories}(g) and~\ref{fig:trajectories}(h) for an in-phase and out-of-phase synchronization, respectively. It is observed that, in this case, the separation distance between the two swimmers (black doted lines) increases, and a side-by-side formation is not achieved. 

The results depicted on Figure~\ref{fig:trajectories} show that a side-by-side formation of two tuna-like bio-robots can be achieved through hydrodynamic interactions between the swimmers. These interactions affect the thrust production of the swimmers creating restorative forces that render the side-by-side formation stable in the streamwise direction. Importantly, these hydrodynamic interactions are \textit{insensitive} to the phase synchronization of the swimmers. This is a sharp deviation from what is observed for schools of oscillating hydrofoils, for which the phase synchronization is the key parameter to dictate the spatial arrangement of stable schooling formations~\citep{kurt2018flow,newbolt2022lateral,ormonde2024two}. The stable side-by-side formation observed for the pair of tuna-like bio-robots also seems to depend on the initial distance between the bio-robots. If the initial separation is sufficiently small, the system is attracted to a stable equilibrium state, while if sufficiently large, no side-by-side or stable formation, in general, is observed. Compared to the experiments, the simulated school achieves the stable equilibrium state in less cycles due to the mass of the simulated bio-robots being lower than in the experiments, so the bio-robots in the simulations experience higher accelerations compared to the experiments. Nevertheless, the stable equilibrium state is the same for both, meaning that this phenomenon is primarily driven by the time-averaged hydrodynamic interactions between swimmers.   

\subsection{Restorative forces leading to the streamwise stability} \label{subsec:hydro_mechanisms}

\begin{figure}
	\centering
	\includegraphics[width=\linewidth]{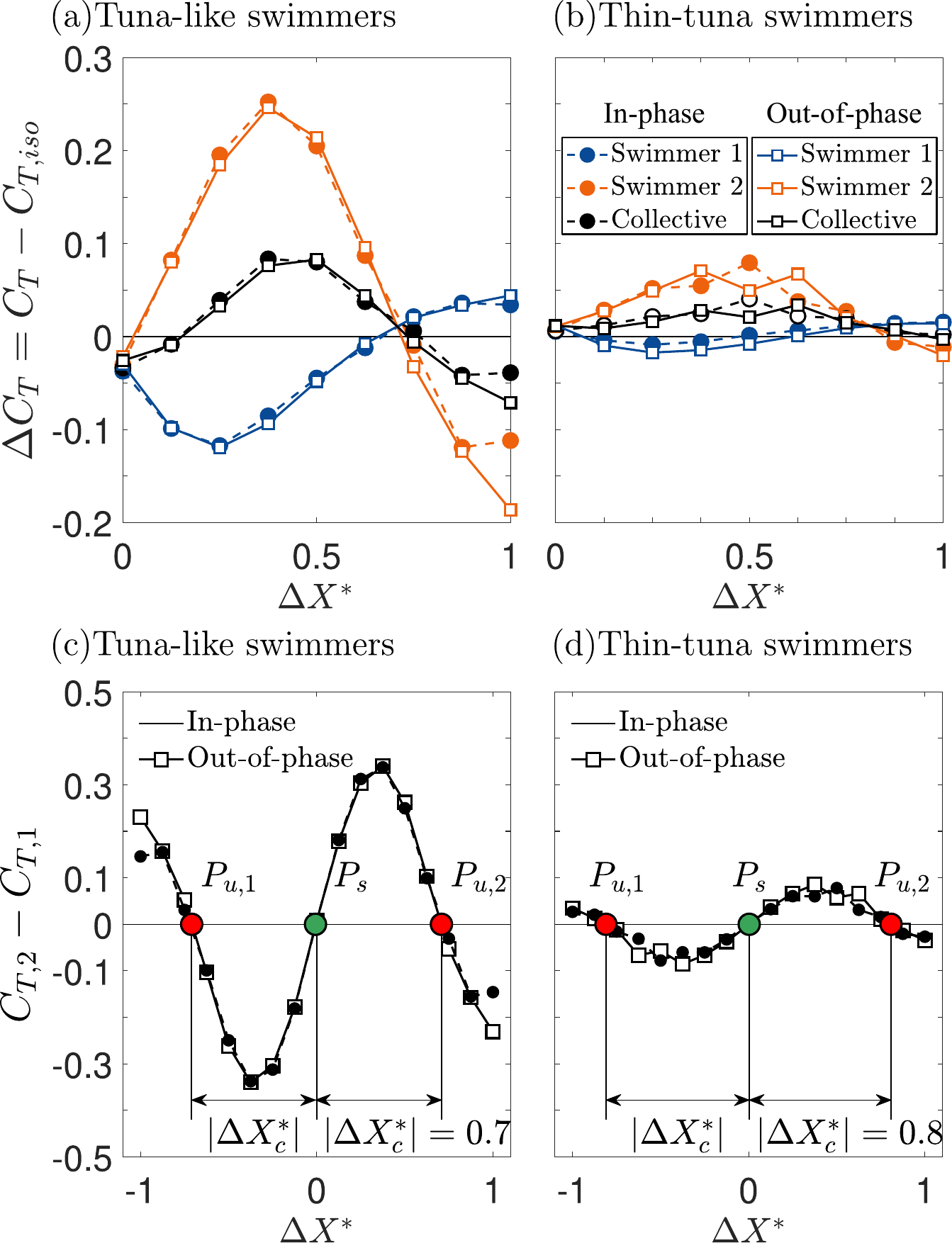}
	\caption{ (a,b) Effect of streamwise spacing on the thrust generation for swimmer 1 ($\Delta C_{T,1}$), swimmer 2 ($\Delta C_{T,2}$) and for the collective ($(\Delta C_{T,1} + \Delta C_{T,2})/2$), from constrained simulations of  (a) tuna-like swimmers, and (b) thin-tuna swimmers. For the tuna-like swimmers, swimmer 1 (blue lines) generates higher thrust compared to swimming in isolation in the region $1/8\leq \Delta X^* \leq 5/8$, while swimmer 2 (orange line) generates lower thrust. Further downstream, at $\Delta X^*\geq7/8$, swimmer 2 has a thrust deficit $\Delta C_{T,1} <0$. The same trends can be observed for the thin-tuna swimmers in (b), albeit at much lower magnitudes. (c,d) Relative thrust $C_{T,2}-C_{T,1}$ as a function of stream-wise spacing for simulations of in-phase and out-of-phase synchronizations for (c) tuna-like swimmers, and (d) thin-tuna swimmers. For the tuna-like swimmers, the relative force between individuals reaches up to $|\Delta C_{T_{2,1}}^{\text{max}}|= 0.34$ at $\Delta X^* = 0.375$ and -0.375. One stable equilibrium point $P_s$ is found for the side-by-side formation (green circle marker), and the two measured unstable equilibrium points $P_{u,1}$ and $P_{u,2}$ (red circle markers) are found for staggered formations $\Delta X^*= -0.7$ and 0.7, respectively, which define the converging distance $|X^*_c| = 0.7$. For the thin-tuna swimmers, the relative changes in thrust are smaller, with $|\Delta C_{T_{2,1}}^{\text{max}}|=0.077.$ A stable equilibrium configuration is found at $\Delta X^* = 0$, as well as two unstable equilibrium points.} 
    \label{fig:forces_constrained}
\end{figure}

Simulations of two \emph{constrained} swimmers were performed to help uncover the mechanisms that generate the restorative forces leading to a stable side-by-side formation. The pair was arranged in a total of nine different streamwise separation distances $\Delta X^* = [0, \ 1/8, \ ..., \ 7/8, \ 1]$ with constant cross-stream spacing $\Delta Y^* = 0.43$. Figure~\ref{fig:forces_constrained}(a) shows the time-averaged thrust of schooling tuna-like swimmers relative to a solitary swimmer, $\Delta C_{T,i} = C_{T,i} - C_{T,\text{iso}}$, as a function of streamwise spacing, where $i=1$ and 2 refers to swimmers 1 and 2, respectively. The average collective thrust variation per swimmer in the school $\Delta C_{T,C} = \left( \Delta C_{T,1} + \Delta C_{T,2} \right)/2$ is also computed. The results show that the downstream swimmer, bio-robot 2, experiences the largest thrust increase of 25\% at $\Delta X^* = 3/5$, with net thrust gains $\Delta C_{T,2}>0$ in the region $1/8\leq X^* \leq 5/8$ compared to an isolated swimmer. The opposite trend occurs for the upstream bio-robot (swimmer 1), experiencing a net thrust deficit $\Delta C_{T,1}<0$ within the same region. Note that the phase synchronization does not affect the trends in the plot. 

Figure~\ref{fig:forces_constrained}(c) shows the time-averaged, net thrust difference between two tuna-like swimmers $\Delta C_{T_{2,1}} = C_{T,2} - C_{T,1}$ as a function of streamwise spacing. $\Delta C_{T_{2,1}}$ is positive if the downstream bio-robot (swimmer 2) produces higher net thrust than the upstream bio-robot (swimmer 1). The green circle marker indicates the location of a \emph{stable equilibrium point} at $\Delta X^* = 0$, where the thrust of both swimmers is the same. The stability of the system is indicated by the positive slope of the curve in the vicinity of the equilibrium point. For $\Delta X^*> 0$ Tuna2 is downstream of Tuna1, and $\Delta C_T > 0$ means that Tuna2 produces a higher average thrust than Tuna1. For $\Delta X^*< 0$, Tuna2 is upstream of Tuna1, and $\Delta C_T < 0$ means that Tuna2 produces less thrust than Tuna1. The net force curves show that the school, if perturbed from a given initial state sufficiently close to the equilibrium state, tends to passively re-organize itself to a side-by-side formation due to the restorative hydrodynamic forces. In other words, the side-by-side formation is stable. Two \emph{unstable} equilibrium points $P_{u,1}$ and $P_{u,2}$ are also found at $\Delta X^* = -0.7$ and 0.7, respectively, and are highlighted by red circles. These unstable points define the converging distance $|X^*_c| = |P_{u,1} - P_s| = |P_{u,2} - P_s|$, within which the relative thrust force $\Delta C_T$ tends to attract the pair to the stable side-by-side formation. For larger separation distances $|X^*| > |X^*_c|$, the time-averaged forces tend to push the pair away from each other, and the side-by-side formation will not be achieved if the kinematics of both swimmers are maintained.

\subsection{Proposed hydrodynamic mechanism behind the stable side-by-side formation}
\label{subsec:bodyshape}

As observed in Figures~\ref{fig:forces_constrained}(a) and~\ref{fig:forces_constrained}(c), the changes in thrust production due to schooling are largely insensitive to the phase synchrony -- the in-phase and out-of-phase cases are almost identical. These results suggest that the changes in the thrust generation of the bio-robots originates from changes in the quasi-steady flow in the vicinity of the bio-robot bodies, and do not depend on time-varying interactions between the wakes shed from the caudal fins and the fins themselves. For schools of oscillatory hydrofoils, the main mechanisms that affect their hydrodynamic performance and formation stability originate from time-varying interactions with their wake structures, which depend on the phase synchronization between the foils~\citep{ramananarivo2016flow,Kurt2018,ormonde2024two}. On the other hand, for the bio-robots in the present study we hypothesize that the changes in their thrust production is caused by the acceleration of flow in the region between the two bio-robots, producing what is called hereafter as a ``channeling effect". Our \textit{hypothesis} is that the flow constriction and acceleration between the tuna-like bio-robots is associated with a time-averaged low pressure zone that acts like a spring between the bio-robots to pull a follower forward and a leader back into a side-by-side formation if perturbed from this equilibrium state. It is not unprecedented to consider that body shape and thickness effects can impact the performance of schooling fish-like swimmers~\citep{KellyDong2024}. 

To test this channeling hypothesis, experiments were devised with thin-tuna swimmers, which are 62\% thinner than the tuna-like body geometry, as depicted in Figure~\ref{fig:coords_thickness}. The reasoning was that if the channeling hypothesis is correct then by reducing the body thickness of the bio-robots, the low-pressure-based restorative forces will get much smaller and perhaps the swimmers will no longer hold a stable formation. Thin-tuna tests will then help us understand the importance of body-body interactions on the streamwise stability of the school with minimal changes to the wake structures produced by the caudal fins of the baseline, tuna-like bio-robots. If this proposed mechanism is accurate then it would rationalize why the side-by-side stable formation is insensitive to the phase synchronization -- that is the steady-flow-based body-body interactions predominately drive the restorative forces.

Indeed, the simulations in Figure~\ref{fig:forces_constrained}(b) show that for all separation distances examined the thrust generation of the thin-tuna bio-robots are not nearly as affected by the interactions with their neighbor as the tuna-like bio-robots. For separations within $1/8 \leq \Delta X^* \leq 5/8$, the downstream bio-robot (swimmer 2), still experiences a higher thrust production compared to swimming in isolation, whereas the upstream bio-robot (swimmer 1) still experiences a lower thrust production. However, the highest variation in the thrust force was only 7.7\% of an isolated swimmer -- a much lower magnitude than the tuna-like bio-robots. As for the relative difference in thrust between two thin-tuna bio-robots, the force map curve in Figure~\ref{fig:forces_constrained}(d) shows the same qualitative behavior from the school of tuna-like bio-robots, but also at much lower magnitudes. A stable point $P_s$ is found at $\Delta X^*=0$, and two unstable points $P_{u,1}$ and $P_{u,2}$ are also observed. 

Considering that the restorative forces behave like a spring to return the swimmers to the side-by-side formation when perturbed, we can model the restorative force as $F_\text{restorative} = \Delta C_{T_{2,1}} = k \Delta X^*$. Then the slope of the curves at the equilibrium stable formations in Figures~\ref{fig:forces_constrained}(c) and \ref{fig:forces_constrained}(d) would be equal to the spring constant, $k = d(\Delta C_{T_{2,1}})/d(\Delta X^*)$, which is a measure of the ``stiffness" or strength of the restorative forces. Under this idea, the thin-tuna swimmers have an over four times weaker stable formation than the thick-tuna swimmers. The spring analogy, while useful for quantifying the strength of the restorative forces, is not a perfect analogy in that the leader and follower bio-robots do not experience equal, but opposite restorative forces as would be the case if a spring connected the two swimmers. In fact, the follower experiences a higher thrust benefit than the magnitude of the leader's thrust penalty (Figure \ref{fig:forces_constrained}). This property of the interaction may prove useful since there is a net thrust benefit for staggered formations around $\Delta X^* = 0.5$.
\begin{figure}[t!]
	\centering
	\includegraphics[width=1\linewidth]{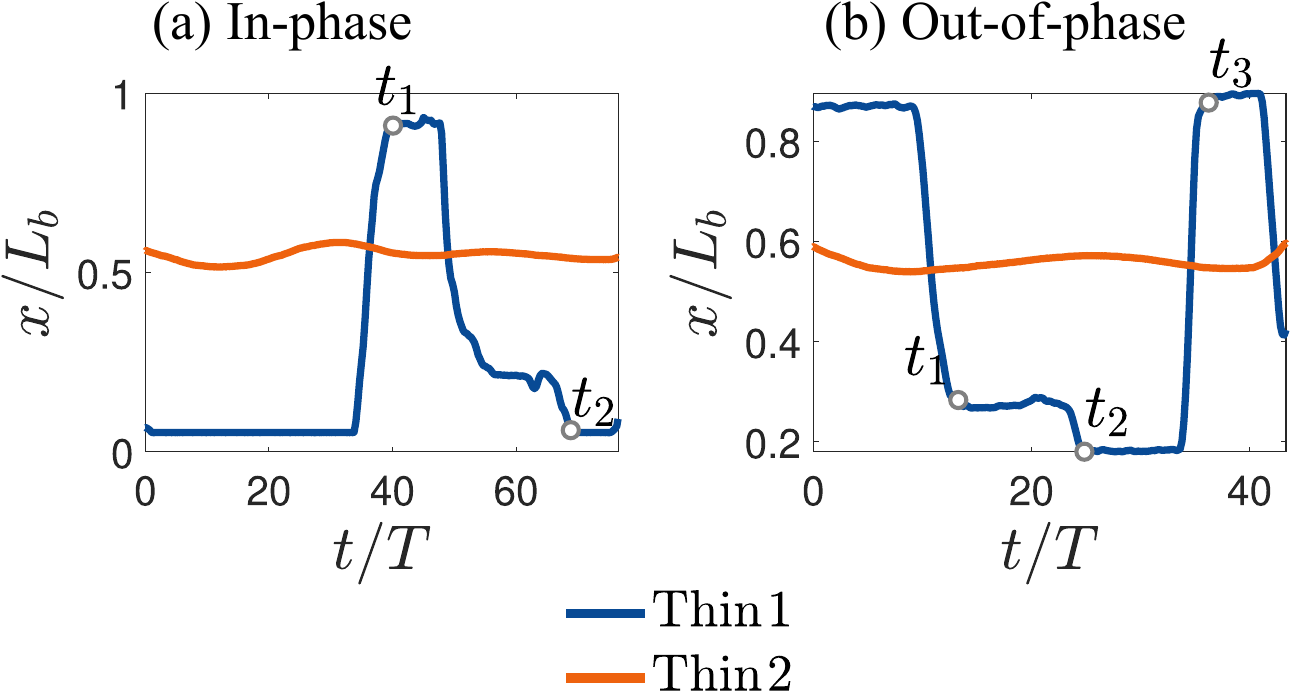}
	\caption{Experimental free-swimming trajectories of swimmer 2 in response to manually prescribed (perturbed) position of swimmer 1, for an (a) in-phase and (b) out-of-phase synchrony. Blue line: swimmer 1; Orange line: swimmer 2; Black line: $\Delta X^*(t)$.}
	\label{fig:thin_trajectories}
\end{figure}

Unconstrained experiments were also performed for the thin-tuna bio-robots where swimmer 2 is free to swim while the location of swimmer 1 is prescribed by manually moving it. Figure~\ref{fig:thin_trajectories}(a) and~\ref{fig:thin_trajectories}(b) shows the trajectories of swimmer 2 in response to the location of swimmer 1, or lack thereof, for an in-phase and out-of-phase synchronization, respectively. In Figure~\ref{fig:thin_trajectories}(a), it can be observed that swimmer 2 (orange line) does not move upstream or downstream in response to the changes in the position of swimmer 1 (blue line) at times $t_1$ and $t_2$, indicating that the thrust of swimmer 2 is roughly constant throughout the experiment. The same behavior is observed on Figure~\ref{fig:thin_trajectories}(b), for which the position of swimmer 2 is virtually unaffected by the different locations of swimmer 1 at times $t_1$, $t_2$, and $t_3$.

It is important to note that the constrained simulations from Figures~\ref{fig:forces_constrained}(c) and~\ref{fig:forces_constrained}(d) indicate the existence of a stable side-by-side formation for both the tuna-like and the thin-tuna body geometries at the specified cross-stream distance $\Delta Y^*=0.43$. However, free-swimming simulations and experiments performed for the thin-tuna bio-robots at multiple initial conditions $\Delta X^*_0$ failed to converge to the side-by-side formations. We attribute this behavior to the fact that the constrained force map does not exactly describe the dynamical behavior of the system constituted by the freely swimming, unconstrained swimmers. It is rather a methodology to identify the time-averaged forces experienced by each individual in a given spatial formation. Since the magnitude of forces experienced by the thin-tuna bio-robots is relatively small, the restorative forces depicted on Figure~\ref{fig:forces_constrained}(d) do not seem to be sufficiently strong to maintain school cohesion. External perturbations, small cruising speed mismatches between the two bio-robots, and fluctuations in the instantaneous thrust production of the bio-robots thus seem to be sufficient to disrupt the side-by-side formation of the thin-tunas when free-swimming.  

These results presented are in agreement with our proposed hypothesis that the primary hydrodynamic mechanism that alters the performance of the school originates from quasi-steady body-body interactions and are not significantly affected by the phase synchronization between the bio-robots. Next, we probe the time-averaged flow and pressure fields to more deeply understand the flow physics of the interacting bio-robots that are affecting their thrust production and, consequently, the streamwise stability of the school. In this way, we will further test the flow and pressure components of our hypothesis.

\subsection{Flow structures and pressure fields}

\begin{figure*}[!ht]
	\centering
	\includegraphics[width=0.85\linewidth]{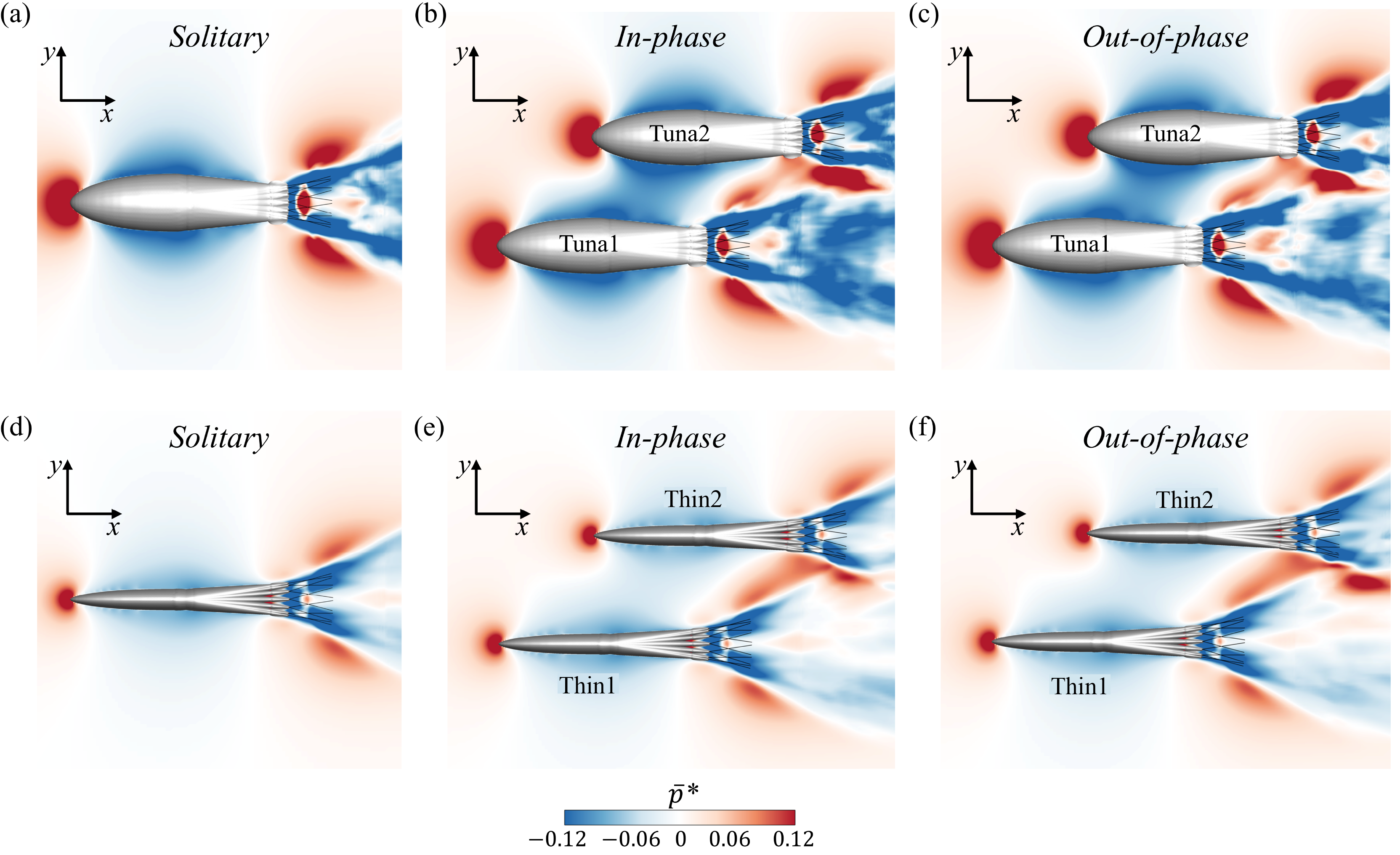}
	\caption{Time-averaged pressure fields at the mid-plane $z=0$. (a) Solitary, tuna-like bio-robot. (b,c) Schooling formation, $\Delta X^* = 3/8$, of tuna-like bio-robots for the in-phase and out-of-phase cases, respectively, showing a decreased pressure in the region between the two swimmers, i.e., the \emph{channeling effect}. (d) Solitary thin-tuna bio-robot. (e,f) Schooling formation, $\Delta X^* = 3/8$, of thin-tuna bio-robots for the in-phase and out-of-phase cases, respectively, showing a less pronounced channeling effect due to the narrower body geometry compared to the tuna-like bio-robot.}
	\label{fig:averaged_pressure_field}
\end{figure*}

\begin{figure*}[!ht]
	\centering
	\includegraphics[width=0.85\linewidth]{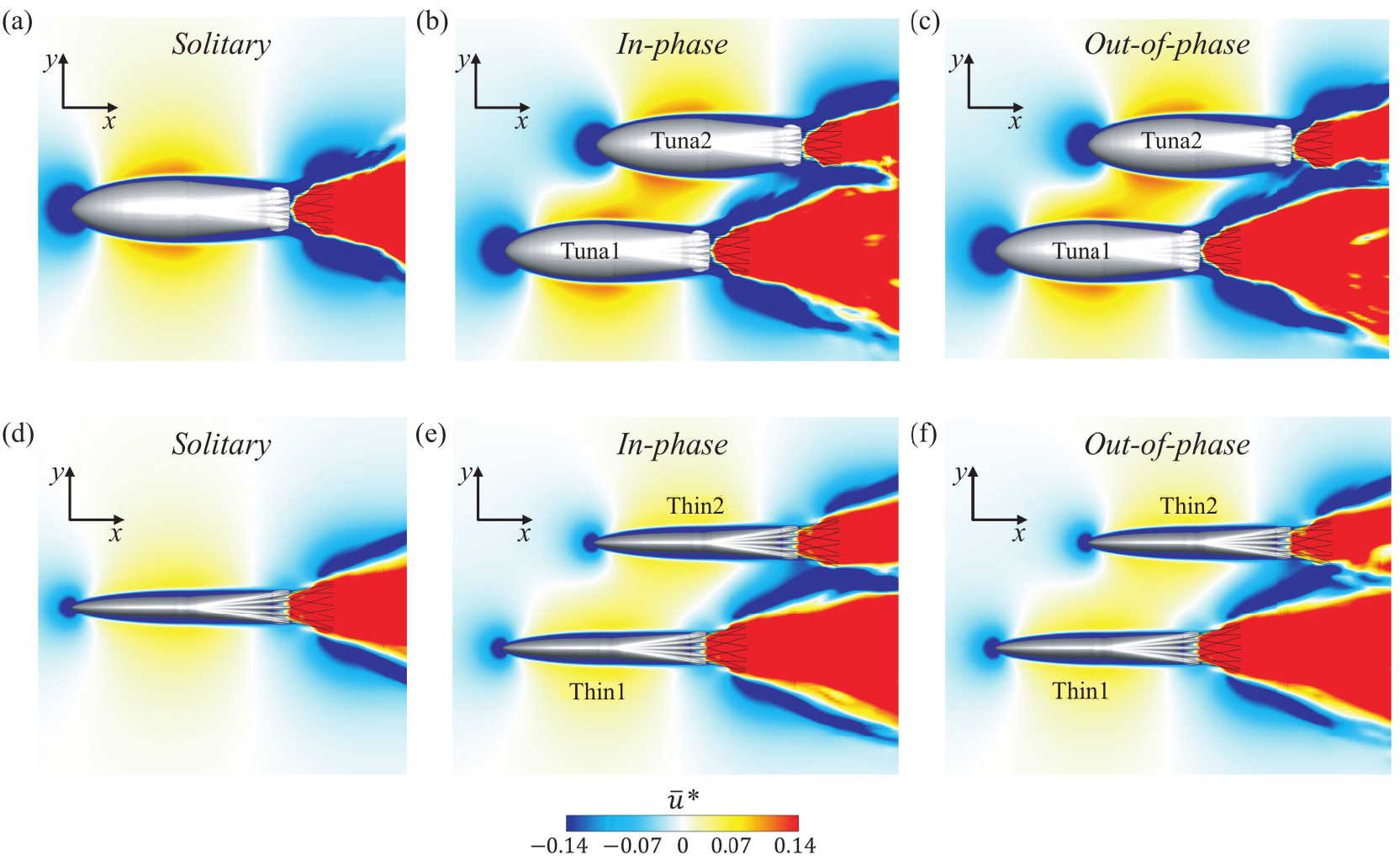}
	\caption{Time-averaged fields of streamwise velocities at the mid-plane $z=0$. (a) Solitary, tuna-like bio-robot. (b,c) Schooling formation, $\Delta X^* = 3/8$, of tuna-like bio-robots for the in-phase and out-of-phase cases, respectively, showing acceleration of flow in the region between the two swimmers, i.e., the \textit{channeling effect}. (d) Solitary thin-tuna bio-robot. (e,f) Schooling formation, $\Delta X^* = 3/8$, of thin-tuna bio-robots for the in-phase and out-of-phase cases, respectively, showing a less pronounced channeling effect due to the narrower body geometry compared to the tuna-like bio-robot.}
	\label{fig:averaged_velocity_field}
\end{figure*}

Figure~\ref{fig:averaged_pressure_field} presents the time-averaged pressure fields for the tuna-like swimmers and thin-tuna swimmers (instantaneous pressure fields can be found in \ref{instantpressure}), and Figure~\ref{fig:averaged_velocity_field} shows the time-averaged velocity fields (instantaneous flow structures can be found in \ref{flowstructures}). All fields are shown for the horizontal plane $z\!=\!0$. Figures~\ref{fig:averaged_pressure_field}(a) and~\ref{fig:averaged_velocity_field}(a) present the solitary tuna-like bio-robot as a reference for the schooling cases. Figures~\ref{fig:averaged_pressure_field}(b) and \ref{fig:averaged_pressure_field}(c) show the schooling tuna-like swimmers in an in-phase and out-of-phase synchronization, respectively, for a staggered formation, $\Delta X^* = 3/8$, which is the formation with highest attractive force $\Delta C_T^*$ between swimmers (see Figure~\ref{fig:forces_constrained}(a)). For both tail synchronizations, a region of low pressure (blue colors) emerges in-between the two bio-robots, indicating a higher flow speed that is confirmed by the streamwise velocity field shown in Figures~\ref{fig:averaged_velocity_field}(b) and \ref{fig:averaged_velocity_field}(c). 

On the left side of the downstream fish (Tuna2), the low pressure region is more prominent at the anterior portion of the body when compared to the solitary swimmer depicted in Figure~\ref{fig:averaged_pressure_field}(a), while the low pressure region is more prominent at the posterior portion of the body of the upstream bio-robot (Tuna1). Moreover, the pressure field around the left side of the tail of Tuna2 is slightly altered as a result of the positively signed pressure produced by the flapping motion of the tail.

\begin{figure}
	\centering
	\includegraphics[width=\linewidth]{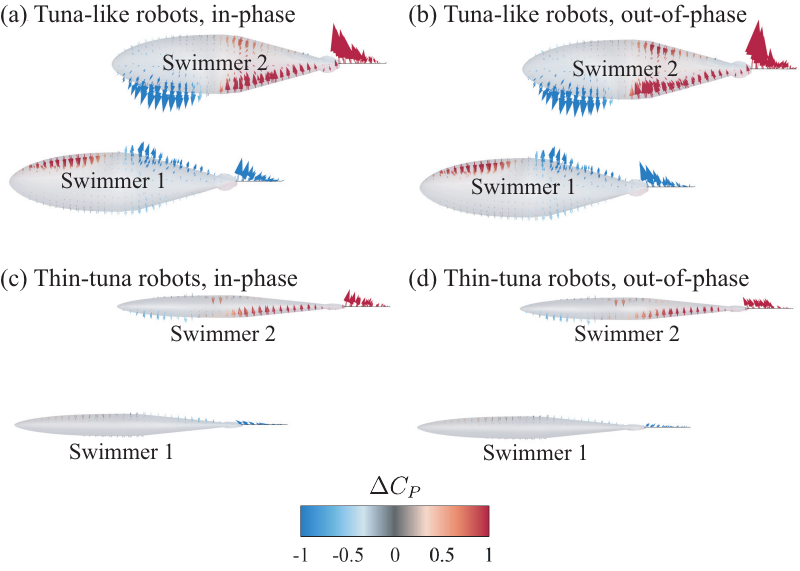}
	\caption{Time-averaged force vectors $\mathbf{\Delta C_F}$ acting on schooling swimmers at $\Delta X^*=3/8$. $\mathbf{\Delta C_F} = \int_A \Delta C_P \vec{n} \mathrm{d}A$ for each mesh element of surface area $A$. Vectors colored with respect to $\Delta C_P = C_P - C_{P,\text{iso}}$. (a) Tuna-like bio-robots, in-phase. (b) Tuna-like bio-robots, out-of-phase. (c) Thin-tuna bio-robots, in-phase. (d) Thin-tuna bio-robots, out-of-phase.}
	\label{fig:forceVecs_dx38}
\end{figure}

Figures~\ref{fig:averaged_pressure_field}(e) and~\ref{fig:averaged_pressure_field}(f) show the same analysis for the pair of thin-tunas at the same formation, $\Delta X^*=3/8$. A low-pressure region in-between the two bio-robots also exists, but with much lower magnitudes in comparison to the pair of tuna-like swimmers, indicating that the flow is not significantly accelerated in that region as observed in Figures~\ref{fig:averaged_velocity_field}(e) and \ref{fig:averaged_velocity_field}(f). The body curvature of the thicker, tuna-like bio-robot acts as a constriction, substantially narrowing the space in between the two swimmers and accelerating the flow within that region, creating the channeling effect. For the pair of thin-tunas, on the other hand, this channeling effect is substantially less pronounced, since their body shape lacks the curvature needed to accelerate the flow.

To understand the connection between these time-averaged pressure field changes and the forces experienced by the swimmers, Figure~\ref{fig:forceVecs_dx38} presents the time-averaged pressure-based force vectors $\boldsymbol{\Delta C_F} = \int_A \Delta C_P \cdot \boldsymbol{n} \mathrm{d}A$ acting on each surface element of area $A$ of the bio-robots for a school formation of $\Delta X^* = 3/8$. Vector lengths represent the relative force magnitudes, and colors represent the change of pressure relative to an isolated swimmer $\Delta C_P = C_P - C_{P,\text{iso}}$. To construct the time-averaged vectors, we take the instantaneous force vectors for each mesh element of the bio-robot in the simulation at each time step of one tail beat cycle, taking into account the posture of the bio-robot at that time step. Finally, the time-averaged vector $\boldsymbol{\Delta C_F}$ of each mesh element is calculated. These vectors are then projected onto the bio-robot with straightened posture, tail and caudal fin parallel to the swimming direction, as a way to visualize the time-average contribution of the surrounding pressure fields acting on each location of the bio-robot's body as depicted on Figure~\ref{fig:forceVecs_dx38}.

For the tuna-like bio-robots in Figures~\ref{fig:forceVecs_dx38}(a) and~\ref{fig:forceVecs_dx38}(b), we observe how the pressure-based force distribution acting on swimmer 2 increases its hydrodynamic performance regardless of the synchronization. The blue force vectors on the anterior portion of the body are pointing in the upstream direction, which indicates an overall pressure reduction on the head. Conversely, the red vectors show a pressure increase in the posterior portion of the body also pointing upstream. Swimmer 2 simultaneously experiences a reduction of pressure drag on its head and an increase in thrust from its propulsive section, leading to an overall increase in the time-averaged net thrust. For swimmer 1, there are red, downstream-pointing forces acting on the head section, indicating an increase in pressure drag whereas the propulsive section produces lower thrust, as indicated by the blue forces pointing slightly downstream. Figures~\ref{fig:forceVecs_dx38}(c) and~\ref{fig:forceVecs_dx38}(d) show the same representation of force vectors for the thin-tuna bio-robots for an in-phase and out-of-phase synchronization, respectively. The same trends obtained for the tuna-like bio-robots are seen for the thin-tuna bio-robots, but at much lower magnitudes. On swimmer 2, the small blue vectors on the head section indicate a reduction in pressure drag, and the red vectors on the propulsive section pointing upstream indicate an increase in thrust from that region. For swimmer 1, no significant changes are observed, as seen by the very small magnitude of the vectors in the figures. Further quantitative assessment of the distribution of the pressure forces can be found in \ref{pressforcedist}.

In this study we find a stable side-by-side formation for bio-robots much like the stable side-by-side formations of oscillatory foils~\citep{ormonde2024two}, however the hydrodynamic mechanism leading to the stable formation of schooling foils depends on the phase synchrony between them; something not observed for the fish-like bio-robots in this study. Now, the time-averaged velocity and pressure fields confirm our hypothesis that the dominant mechanism driving the stable formation of bio-robots is indeed due to flow acceleration in the region between the swimmers from their body curvature, i.e. the channeling effect. Since this is a quasi-steady effect, it is independent of the phase synchrony, which makes its implementation in a bio-robotic school relatively simple. For instance, to promote or avoid the channeling effect between fish-like bio-robots, one needs to take into consideration the relative thickness of the bio-robots and their lateral spacing, but does not need to take into consideration the phase synchronization or even the flapping frequency of the propulsors, since the hydrodynamic mechanism is largely insensitive to the time-varying flowfield. The results found in the present study thus provide new control opportunities for the promotion and maintenance of group cohesion of schools of bio-robots that may also be occurring in real fish schools.

\subsection{Hydrodynamic performance}
\label{sec:performance}
The performance of schooling formations of the two bio-robot body geometries is compared to the performance of swimming in isolation. Performance metrics of unstable (staggered) school formations are obtained from constrained simulations, and the performance of the stable side-by-side formations of the tuna-like bio-robots is measured from unconstrained simulations and experiments.

\begin{figure}
    \centering
    \includegraphics[width=\linewidth]{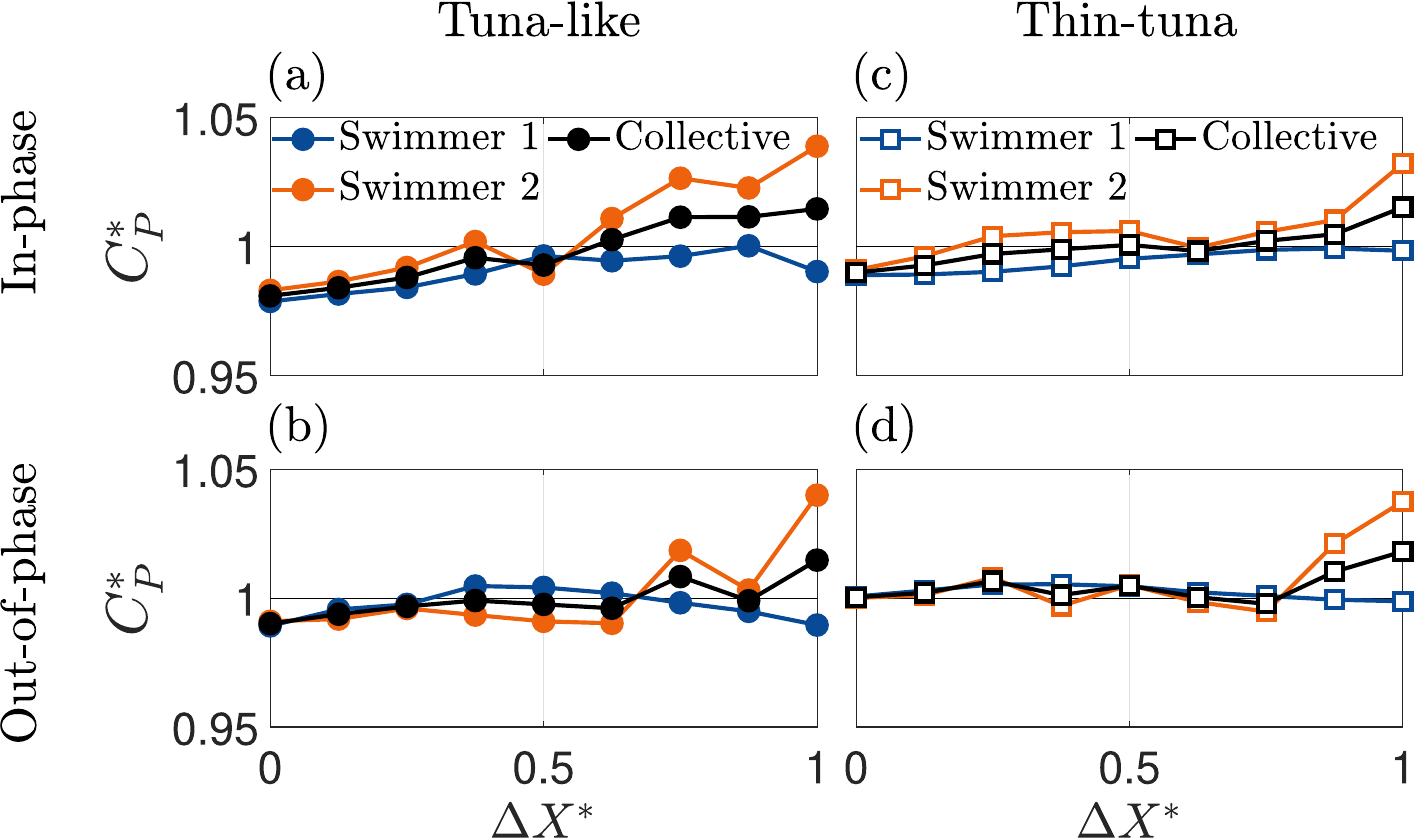}
    \caption{Power coefficient for the constrained simulations. (a,b) Tuna-like swimmers for an in-phase and out-of-phase synchronization, respectively. (c,d) Thin-tuna swimmers for an in-phase and out-of-phase synchronization, respectively. For all cases, the collective power (black line) increases with the streamwise separation distance, with the power of the upstream, swimmer 1 (blue line) remaining within 2\% of the isolated case, while the downstream, swimmer 2 (orange dots) increases with $\Delta X^*$ for all cases.}
    \label{fig:Cp_constrained}
\end{figure}

Figure~\ref{fig:Cp_constrained} presents the collective power coefficient of the constrained cases previously discussed in Figure~\ref{fig:forces_constrained}, where swimmer 2 is placed at fixed locations of $\Delta X^* = [0, \ 1/8, \ ..., \ 7/8, \ 1]$. The power coefficients of swimmer 1 (blue line), swimmer 2 (orange line), and for the collective (black line) are normalized by that of an isolated swimmer $C_P^* = C_P/C_{P,\text{iso}}$. For all four cases, swimmer 2 experiences an increase of 4\% in power expenditure when located at $\Delta X^* = 1$, whereas swimmer 1 has variations of less than 2\% in $C_P^*$ for all formations. The constrained simulations allow us to understand the general performance trends resulting from the formations of each swimmer. 

The time-averaged performance of the tuna-like bio-robots in the stable side-by-side formation is presented in Table~\ref{tab:freeswim_perf}. The final swimming speed of the school is lower than that of an isolated swimmer $U_{\text{iso}}$ in both the experiments and simulations. Compared to an isolated swimmer, the collective swimming speed is 4\% lower in the experiments and 2\% lower in the simulations; the collective power coefficient is 11\% higher in the experiments and 4\% higher in the simulations, and the cost of transport is 4\% higher in the experiments and remains effectively unaltered in the simulations. Overall, these performance metrics are insensitive to the tail beat synchronization, since the wake structures from the two swimmers do not interact significantly with the swimmers as observed in Figure~\ref{fig:IP-regbody-flowfield_topview}. 

The collective performance of the tuna-like bio-robots in the stable side-by-side formation indicate a small reduction in thrust and swimming speed, and increased power expenditure regardless of their phase synchrony. By contrast, the thrust and power of oscillatory foils in a side-by-side formation depends on the phase synchrony. An in-phase synchronization reduces thrust production and power expenditure, while an out-of-phase synchrony increases thrust and power~\citep{ormonde2024two,kurt2019swimming,Boschitsch2014}. Since the hydrodynamic mechanism underlying the stable side-by-side formation of foils differs from that of fish-like bio-robots, the effect of schooling on their collective performance is also different. Foils experience changes in their circulatory forces~\citep{han2023revealing} due to vortex-body interactions that depend upon their synchrony. The performance of the bio-robots in the present study is primarily affected by the quasi-steady effect of flow acceleration between the swimmers, which is independent from their tails' synchronization.  
\begin{table}[h]
\centering
\resizebox{\columnwidth}{!}{%
\begin{tabular}{c c c|c c|c c}
\multirow{2}{*}{} & \multicolumn{2}{c|}{$U^*$} & \multicolumn{2}{c|}{$C_{P,C}^*$} & \multicolumn{2}{c}{$CoT^*$} \\
 & exp. & sim. & exp. & sim. & exp. & sim. \\
 \cline{2-7}
In-phase & 0.96 & 0.981 & 1.11& 1.035 & 1.04& 0.995\\
Out-of-phase & 0.96 & 0.982 & 1.11& 1.041 & 1.04& 1.003\\
\end{tabular}%
}
\caption{Collective swimming speed, power coefficient, and cost of transport  measured for the stable side-by-side formation of freely swimming tuna-like swimmers.}
\label{tab:freeswim_perf}
\end{table}

\section{Conclusions and discussion}
\label{sec:conclusions}
In this study, we advance our understanding of schools of three-dimensional, fish-like swimmers by allowing them to swim freely in one degree-of-freedom in the streamwise direction in both simulations and experiments.  We show, for the first time, that independently free-swimming tuna-like bio-robots experience a fluid-mediated stable side-by-side formation. Specifically, we reveal that the body thickness of fish-like swimmers affects the hydrodynamic interactions that occur between bio-robots when the flow between the bodies is accelerated in a ``channeling effect'' due to flow constriction. Consequently, this creates a low-pressure zone that is the primary mechanism driving a fluid-mediated restorative force around a side-by-side formation, thereby making it a hydrodynamically stable formation. This quasi-steady channeling mechanism makes the stability of the side-by-side formation insensitive to the phase synchronization between the fins of the bio-robots in contrast to previous results of stable side-by-side formations observed for oscillatory foils, for which the phase synchronization and vortex-body interactions are the main mechanisms behind the stable formation~\citep{ormonde2024two}. 

For a pair of thick-bodied (maximum thickness of 22\% of the total length) tuna-like swimmers placed in a staggered formation, the leading swimmer perceives a net increase in pressure drag, thus decelerating the swimmer. The trailing swimmer, on the other hand, perceives a net reduction in pressure drag causing it to accelerate forwards, towards the decelerating upstream swimmer. From constrained-swimming simulations, the effect of those relative forces is to attract the pair of tuna-like swimmers together with a peak net thrust of $\Delta C_T = 0.34$ when the streamwise separation is $\Delta X^*=0.375$ and cross-stream separation is $\Delta Y^*=0.43$. For the more slender bio-robots (``thin-tunas'') with a maximum body thickness of 8\% of their length, this effect is also present, but greatly diminished. For the thin-tuna swimmers the flow acceleration between the two bodies is less pronounced, leading to relative attractive force that only reaches a peak value of 22\% of the peak net thrust of the thicker, tuna-like swimmers. As a consequence, in free-swimming simulations and experiments a stable side-by-side formation was \textit{only} observed for the pair of thick-bodied, tuna-like swimmers. This indicates that the time-averaged hydrodynamic effects found in constrained-swimming simulations are a good indicator of schooling effects if the mean forces are sufficiently large to dictate the dynamic behavior of the school.

When swimming in the stable side-by-side formation, the tuna-like swimmers are seen to have a small reduction in their swimming speed and a concurrent small rise in their cost of transport. That is, the side-by-side formation may be stable, but it is not providing an energetic or speed benefit in the current implementation. However, considering other strategies \citep{Han_2025} in conjunction with this stable formation may be leveraged to promote energetic savings while maintaining a modified stable formation. What's clear is this study reveals an opportunity to design the body shape of bio-robots that promote the channeling effect, which can then be exploited to offload some of the demands on real-time control systems required to maintain group cohesion. This strategy may be more simple and robust than relying on, for instance, wake-body interactions that depend on the precise control of tail beat frequency and phase synchronization, since the channeling effect is insensitive to variations in the frequency or synchronization of the swimmers.  

\appendix
\section{Instantaneous pressure fields for side-by-side and staggered formations}\label{instantpressure}

\begin{figure*}[!ht]
	\centering
	\includegraphics[width=\linewidth]{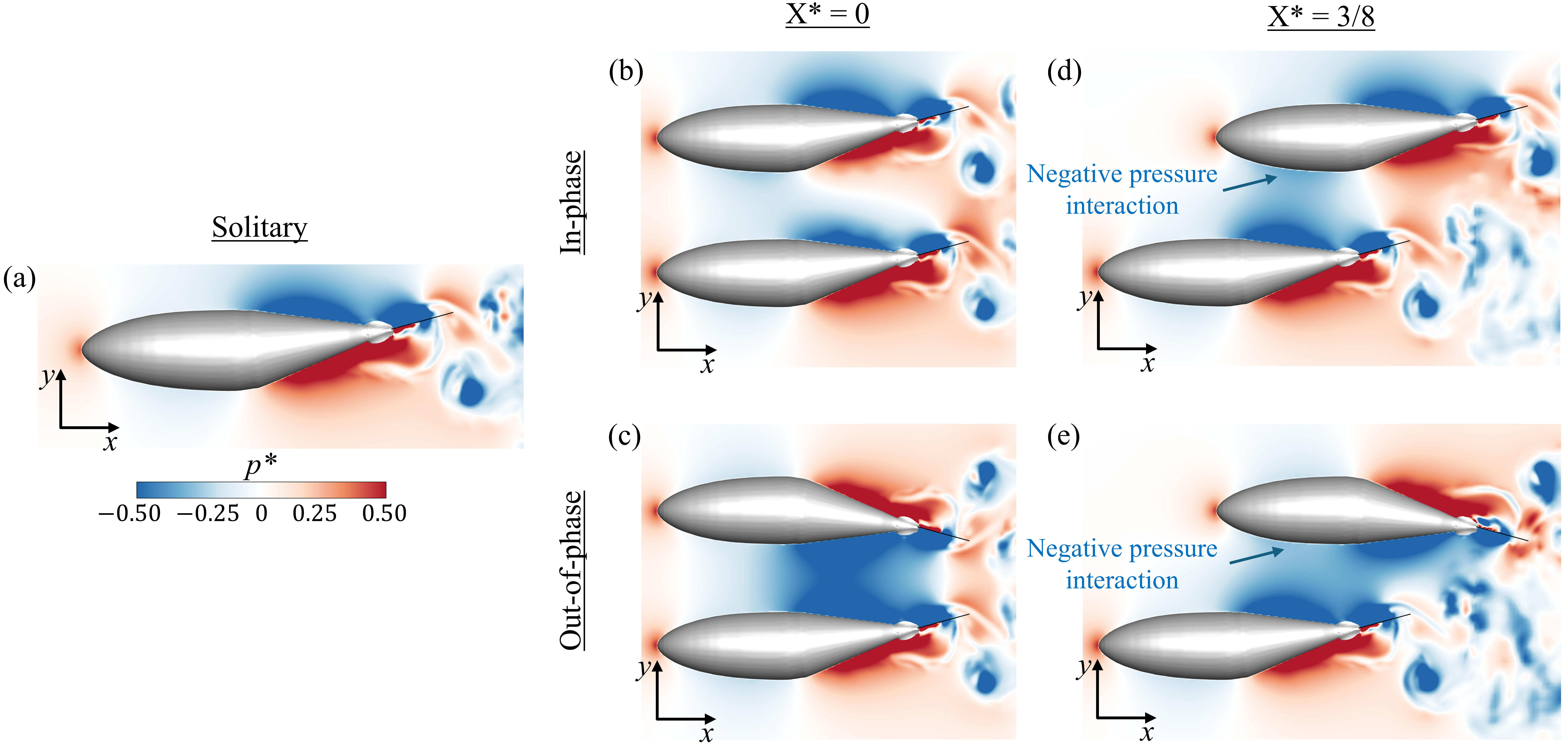}
	\caption{Slice cuts of the pressure field taken at the mid-plane of the tuna-like swimmers at $t/T = 5.0$. (a) Solitary swimmer for reference. $X^* = 0$ is shown in (b,c) for both the in-phase and out-of-phase kinematics. $X^* = 3/8$ is shown in (d,e) for both the in-phase and out-of-phase kinematics.}
	\label{fig:IPOP-regbody-presure_topview}
\end{figure*}

\begin{figure*}[!ht]
	\centering
	\includegraphics[width=1\linewidth]{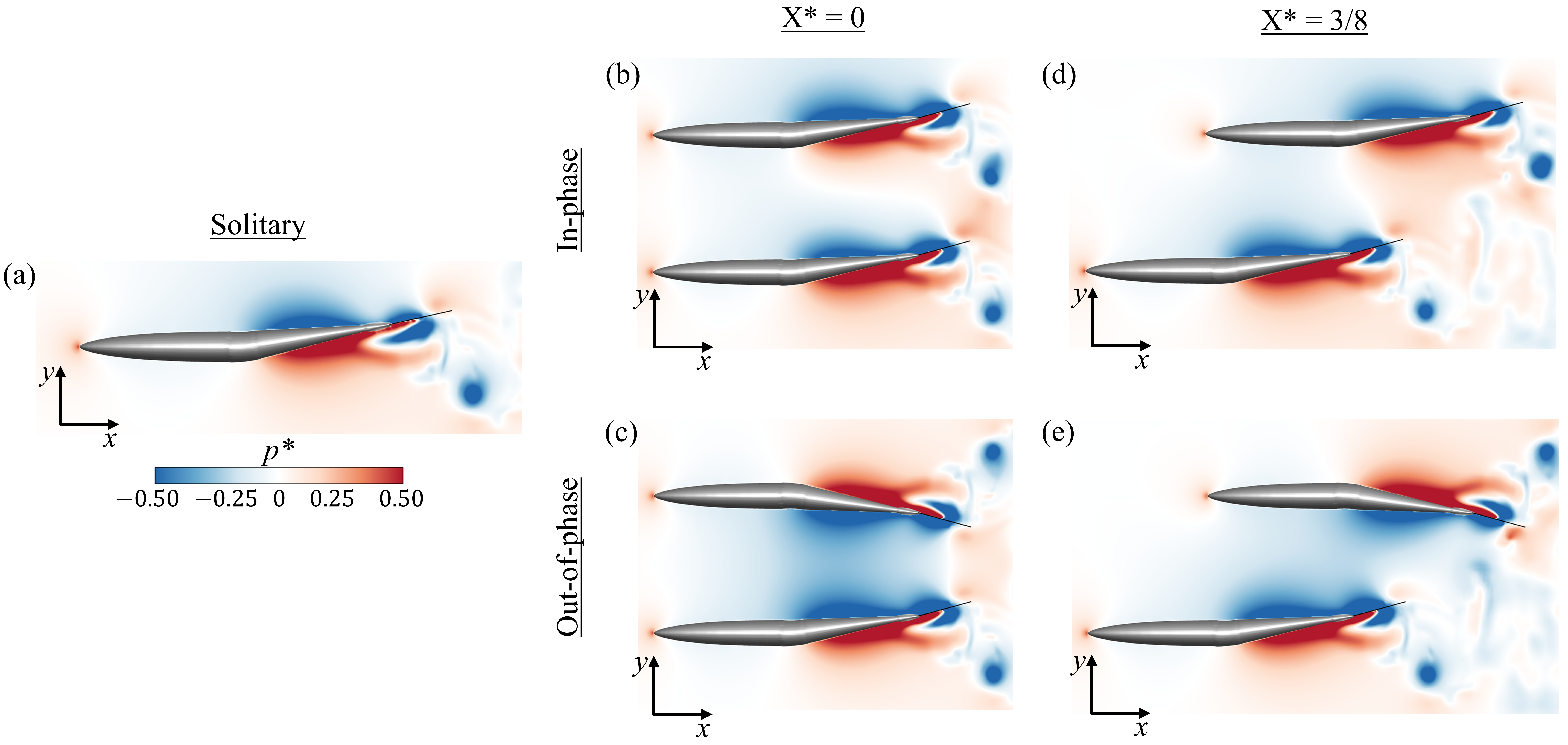}
	\caption{Slice cuts of the pressure field taken at the mid-plane of the thin-tuna swimmers at $t/T = 5.0$. (a) Solitary swimmer for reference. $X^* = 0$ is shown in (b,c) for both the in-phase and out-of-phase kinematics. $X^* = 3/8$ is shown in (d,e) for both the in-phase and out-of-phase kinematics.}
	\label{fig:IPOP-thinbody-presure_topview}
\end{figure*}

\section{Flow structures and caudal fin}\label{flowstructures}
\begin{figure*}[!ht]
	\centering
	\includegraphics[width=1\linewidth]{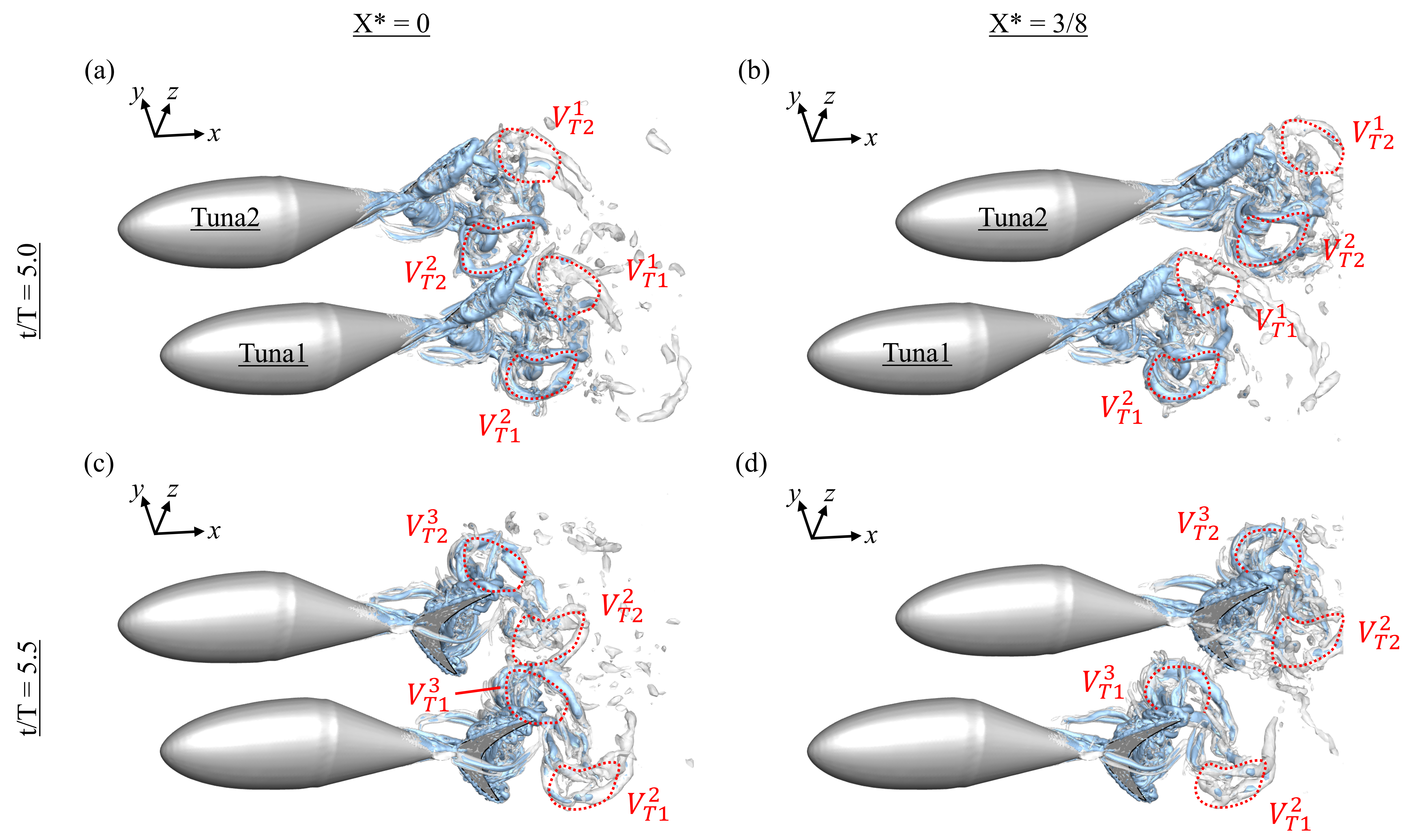}
	\caption{Q-criterion wake visualization for $X* = 0$ (a,c) and $X* = 3/8$ (b,d). Wakes are shown at two time instances during the flapping motion; one at the end of the right stroke ($t/T = 5.0$) in (a,b) and another at the end of the left stroke ($t/T = 5.5$) in (c,d).}
	\label{fig:IP-regbody-flowfield_topview}
\end{figure*}

Figure~\ref{fig:IP-regbody-flowfield_topview} shows the flow field for two tuna-like swimmers schooling in-phase at two spacings: $X^* = 0$ and $3/8$. The Q-criterion is visualized at two iso-surfaces: $Q = 150$ (blue) and $Q = 50$ (transparent gray). Figures~\ref{fig:IP-regbody-flowfield_topview}(a) and (b) show wake structures at the end of a rightwards flapping motion (rotation in the $-z$ direction) at time $t/T=5.0$. 

The overall arrangement of the wake structures at $t/T=5.0$ in Figures~\ref{fig:IP-regbody-flowfield_topview}(a) and \ref{fig:IP-regbody-flowfield_topview}(b) provide an understanding of the relative distances between the wakes of Tuna1 and Tuna2. Both tunas have just completed a rightward flapping motion. $V^{1}_{T1}$ (labeled with a subscript according to the fish from which the vortex generated, and a superscript according to the order in which the vortex ring was generated with 1 being the oldest ring) is seen approaching Tuna2 in $X^* = 3/8$, however, the interaction between the wakes of one swimmer and the body/fin of another is non-existent for $X^* = 0$ given that the caudal fins of both Tuna1 and Tuna2 are located at the same streamwise position. Coherent vortex structures are seen forming on the caudal fins of both Tuna1 and Tuna2 as well. At $t/T = 0.5$ in Figure~\ref{fig:IP-regbody-flowfield_topview}(c,d), vortex loop $V^{1}_{T1}$ from Tuna1 dissipates rapidly as it advects downstream, hence, minimal wake interaction between Tuna1 and Tuna2 occurs in this formation. The coherent vortex loop, $V^{3}_{T1}$, observed on the caudal fin in the previous time instance (Figure~\ref{fig:IP-regbody-flowfield_topview}(a,b)), also approaches Tuna2. However, given the periodic nature of the kinematics, this structure will dissipate like seen with $V^{1}_{T1}$ and yield no substantial interaction.

\section{Streamwise distribution of pressure-based forces}\label{pressforcedist}
\begin{figure}
	\centering
	\includegraphics[width=\linewidth]{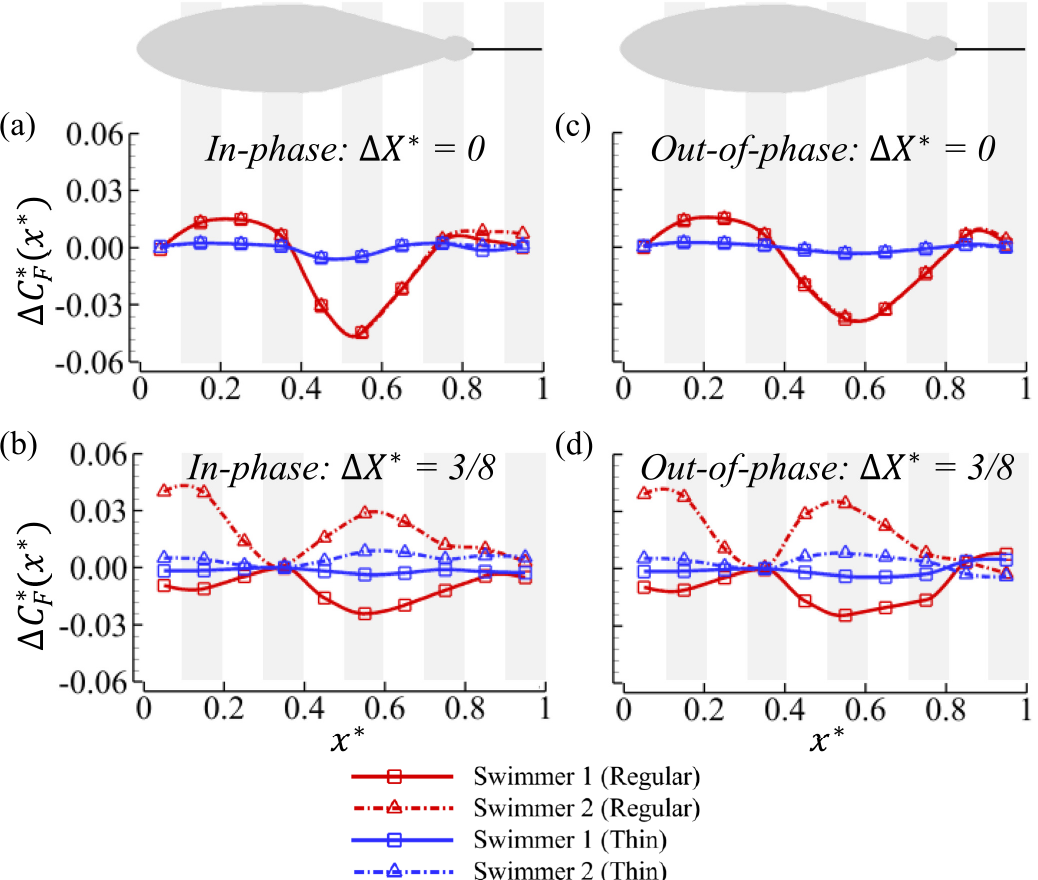}
	\caption{Distribution of pressure-based force in the streamwise direction for schooling bio-robots relative to a solitary swimmer $\Delta C^*_F(x) = \left( C_F(x) - C^{\text{iso}}_F(x)\right)/C^{\text{iso}}_F(x)$. (a) $\Delta X^*\!= \!0$, in-phase. (b) $\Delta X^*\!= \!0$, out-of-phase. (c) $\Delta X^*\!= \!3/8$, in-phase. (d) $\Delta X^* \!= \!3/8$, out-of-phase.}
	\label{fig:body_force_comp}
\end{figure}
Figure~\ref{fig:body_force_comp} provides a quantitative assessment of the distribution of pressure-based force as a function of streamwise location along the body of the tuna-like swimmers. The pressure acting along the surface of the schooling bio-robots relative to the isolated swimmer $\Delta C_p(x) = C_P(x) - C_{P,\text{iso}}(x)$ is used to calculate the streamwise component of the force coefficient $C^*_{F}(x){\ihat} = \int_A \Delta C_p(x) \cdot \boldsymbol{n_{\ihat}} \mathrm{d}A$, where $\boldsymbol{n_{\ihat}}$ is the unit vector in the streamwise direction and $\mathrm{d}A$ is the differential element of the surface area $A$. For the side-by-side formation ($\Delta X^*\!=\!0$) shown by Figures~\ref{fig:body_force_comp}(a) and~\ref{fig:body_force_comp}(c), the pressure-based force distribution for swimmer 1 (red line) is almost identical to that of swimmer 2 (red dotted line) due to the symmetry of the formation, irrespective of the phase synchronization. As a result, the hydrodynamic performance of both swimmers is the same. The slight difference between the curves for in-phase (Figure~\ref{fig:body_force_comp}(a)) and out-of-phase (Figure~\ref{fig:body_force_comp}(b)) synchronizations is caused by the hydrodynamic interactions between the caudal fins and the shed wake structures. These fin-fin interactions are not sufficiently strong to change the topological features of the stable equilibrium point $\Delta X^* = 0$ as discussed in Section~\ref{subsec:hydro_mechanisms}. For the staggered formation, Figures~\ref{fig:body_force_comp}(b) and~\ref{fig:body_force_comp}(c) clearly indicate how the pressure-based force distributions along swimmer 1 and swimmer 2 are affected by the presence of their neighbor. The upstream swimmer (solid red line) experiences an overall decrease in the pressure-based force relative to an isolated swimmer. This is indicated by $\Delta C^*_F(x)<0$ along the entire body length, i.e. a net drag force. Conversely, the downstream swimmer (dotted red line)  experiences an overall increase in the pressure-based force along the entire body, as indicated by $\Delta C^*_F(x)>0$, i.e. a net thrust force. The force distribution curves for the schools of thin-tuna swimmers are depicted by the blue lines in Figure~\ref{fig:body_force_comp}. In all four cases the measured effects caused by the neighbor are small, in agreement with the flow and pressure field results presented in Figures~\ref{fig:IPOP-thinbody-presure_topview}, and~\ref{fig:averaged_pressure_field}.

\clearpage

\bibliographystyle{apalike}
\bibliography{references}

@article{lauder2022robotics,
  title={Robotics as a comparative method in ecology and evolutionary biology},
  author={Lauder, George V},
  journal={Integrative and Comparative Biology},
  volume={62},
  number={3},
  pages={721--734},
  year={2022},
  publisher={Oxford University Press}
}

@article{gravish2018robotics,
  title={Robotics-inspired biology},
  author={Gravish, Nick and Lauder, George V},
  journal={Journal of Experimental Biology},
  volume={221},
  number={7},
  pages={jeb138438},
  year={2018},
  publisher={The Company of Biologists Ltd}
}

@article{ormonde2024two,
	title={Two-dimensionally stable self-organisation arises in simple schooling swimmers through hydrodynamic interactions},
	author={Ormonde, Pedro C and Kurt, Melike and Mivehchi, Amin and Moored, Keith W},
	journal={Journal of Fluid Mechanics},
	volume={1000},
	pages={A90},
	year={2024},
	publisher={Cambridge University Press}
}

@article{Burgerhout2013,
  title={Schooling reduces energy consumption in swimming male European eels, Anguilla anguilla L.},
  author={Burgerhout, Erik and Tudorache, Christian and Brittijn, Sebastiaan A and Palstra, Arjan P and Dirks, Ron P and van den Thillart, Guido EEJM},
  journal={Journal of experimental marine biology and ecology},
  volume={448},
  pages={66--71},
  year={2013},
  publisher={Elsevier}
}

@article{Kurt2018,
  title={Flow interactions of two-and three-dimensional networked bio-inspired control elements in an in-line arrangement},
  author={Kurt, Melike and Moored, Keith W},
  journal={Bioinspiration \& biomimetics},
  volume={13},
  number={4},
  pages={045002},
  year={2018},
  publisher={IOP Publishing}
}

@misc{han2023revealing,
      title={Revealing the mechanism and scaling laws behind equilibrium altitudes of near-ground pitching hydrofoils}, 
      author={Tianjun Han and Qiang Zhong and Amin Mivehchi and Daniel B. Quinn and Keith W. Moored},
      year={2023},
      eprint={2304.14562},
      archivePrefix={arXiv},
      primaryClass={physics.flu-dyn}
}

@article{Boschitsch2014,
author = {Boschitsch, Birgitt M. and Dewey, Peter a. and Smits, Alexander J.},
doi = {10.1063/1.4872308},
issn = {1070-6631},
journal = {Physics of Fluids},
month = {may},
number = {5},
pages = {051901},
title = {{Propulsive performance of unsteady tandem hydrofoils in an in-line configuration}},
url = {http://scitation.aip.org/content/aip/journal/pof2/26/5/10.1063/1.4872308},
volume = {26},
year = {2014}
}

@article{Kurt2020,
author = {Kurt, M. and Panah, A. E. and Moored, K. W.},
doi = {10.3390/biomimetics5020013},
issn = {2313-7673},
journal = {Biomimetics},
number = {2},
pages = {13},
title = {{Flow interactions between low aspect ratio hydrofoils in In-line and staggered arrangements}},
volume = {5},
year = {2020}
}

@article{kurt2019swimming,
  title={Swimming freely near the ground leads to flow-mediated equilibrium altitudes},
  author={Kurt, Melike and Cochran-Carney, Jackson and Zhong, Qiang and Mivehchi, Amin and Quinn, Daniel B and Moored, Keith W},
  journal={Journal of Fluid Mechanics},
  volume={875},
  year={2019},
  publisher={Cambridge University Press}
}

@article{kurt2018flow,
  title={Flow interactions of two-and three-dimensional networked bio-inspired control elements in an in-line arrangement},
  author={Kurt, Melike and Moored, Keith W},
  journal={Bioinspiration \& biomimetics},
  volume={13},
  number={4},
  pages={045002},
  year={2018},
  publisher={IOP Publishing}
}

@article{moored2019inviscid,
  title={Inviscid scaling laws of a self-propelled pitching airfoil},
  author={Moored, Keith W and Quinn, Daniel B},
  journal={AIAA Journal},
  volume={57},
  number={9},
  pages={3686--3700},
  year={2019},
  publisher={American Institute of Aeronautics and Astronautics}
}

@article{weihs1973hydromechanics,
  title={Hydromechanics of fish schooling},
  author={Weihs, D},
  journal={Nature},
  volume={241},
  number={5387},
  pages={290--291},
  year={1973},
  publisher={Nature Publishing Group}
}

@book{lighthill1975mathematical,
  title={Mathematical biofluiddynamics},
  author={Lighthill, Sir James},
  year={1975},
  publisher={SIAM}
}

@article{becker2015hydrodynamic,
  title={Hydrodynamic schooling of flapping swimmers},
  author={Becker, Alexander D and Masoud, Hassan and Newbolt, Joel W and Shelley, Michael and Ristroph, Leif},
  journal={Nature communications},
  volume={6},
  number={1},
  pages={1--8},
  year={2015},
  publisher={Nature Publishing Group}
}

@article{ramananarivo2016flow,
  title={Flow interactions lead to orderly formations of flapping wings in forward flight},
  author={Ramananarivo, Sophie and Fang, Fang and Oza, Anand and Zhang, Jun and Ristroph, Leif},
  journal={Physical Review Fluids},
  volume={1},
  number={7},
  pages={071201},
  year={2016},
  publisher={APS}
}

@article{smits2019undulatory,
  title={Undulatory and oscillatory swimming},
  author={Smits, Alexander J},
  journal={Journal of Fluid Mechanics},
  volume={874},
  pages={P1},
  year={2019},
  publisher={Cambridge University Press}
}

@article{king2018experimental,
  title={Experimental observations of the three-dimensional wake structures and dynamics generated by a rigid, bioinspired pitching panel},
  author={King, Justin T and Kumar, Rajeev and Green, Melissa A},
  journal={Physical Review Fluids},
  volume={3},
  number={3},
  pages={034701},
  year={2018},
  publisher={APS}
}

@article{buchholz2008wake,
  title={The wake structure and thrust performance of a rigid low-aspect-ratio pitching panel},
  author={Buchholz, James HJ and Smits, Alexander J},
  journal={Journal of fluid mechanics},
  volume={603},
  pages={331--365},
  year={2008},
  publisher={Cambridge University Press}
}

@article{parker1973reduced,
  title={Reduced metabolic rates in fishes as a result of induced schooling},
  author={Parker Jr, Faust R},
  journal={Transactions of the American Fisheries Society},
  volume={102},
  number={1},
  pages={125--131},
  year={1973},
  publisher={Wiley Online Library}
}

@article{kelly2023hydrodynamics,
  title={Hydrodynamics of body--body interactions in dense synchronous elongated fish schools},
  author={Kelly, John and Pan, Yu and Menzer, Alec and Dong, Haibo},
  journal={Physics of Fluids},
  volume={35},
  number={4},
  year={2023},
  publisher={AIP Publishing}
}

@article{thandiackal2023line,
  title={In-line swimming dynamics revealed by fish interacting with a robotic mechanism},
  author={Thandiackal, Robin and Lauder, George},
  journal={elife},
  volume={12},
  pages={e81392},
  year={2023},
  publisher={eLife Sciences Publications Limited}
}

@article{zhang2023energetics,
  title={Energetics of collective movement in vertebrates},
  author={Zhang, Yangfan and Lauder, George V},
  journal={Journal of Experimental Biology},
  volume={226},
  number={20},
  pages={jeb245617},
  year={2023},
  publisher={The Company of Biologists Ltd}
}

@article{baddoo2023generalization,
  title={Generalization of waving-plate theory to multiple interacting swimmers},
  author={Baddoo, Peter J and Moore, Nicholas J and Oza, Anand U and Crowdy, Darren G},
  journal={Communications on Pure and Applied Mathematics},
  volume={76},
  number={12},
  pages={3811--3851},
  year={2023},
  publisher={Wiley Online Library}
}

@article{Pan2024computational,
author = {Pan, Yu and Zhang, Wei and Kelly, John and Dong, Haibo},
doi = {10.1063/5.0201965},
journal = {Physics of Fluids},
publisher = {AIP Publishing LLC},
title = {{Unraveling hydrodynamic interactions in fish schools: A three-dimensional computational study of in-line and side-by-side configurations}},
volume = {36},
year = {2024}
}

@inproceedings{pan2022computational,
  title={Computational modeling and hydrodynamic analysis of fish schools in three-dimensional arrangements},
  author={Pan, Yu and Zhang, Wei and Dong, Haibo},
  booktitle={Fluids Engineering Division Summer Meeting},
  volume={85840},
  pages={V002T05A024},
  year={2022},
  organization={American Society of Mechanical Engineers}
}

@article{newbolt2022lateral,
  title={Lateral flow interactions enhance speed and stabilize formations of flapping swimmers},
  author={Newbolt, Joel W and Zhang, Jun and Ristroph, Leif},
  journal={Physical Review Fluids},
  volume={7},
  number={6},
  pages={L061101},
  year={2022},
  publisher={APS}
}

@article{KellyDong2024,
    author = {Kelly, John and Dong, Haibo},
    title = {Effects of body shape on hydrodynamic interactions in a dense diamond fish school},
    journal = {Physics of Fluids},
    volume = {36},
    number = {3},
    pages = {031907},
    year = {2024},
    month = {03},
    issn = {1070-6631},
    doi = {10.1063/5.0192463},
    url = {https://doi.org/10.1063/5.0192463},
    eprint = {https://pubs.aip.org/aip/pof/article-pdf/doi/10.1063/5.0192463/19727896/031907\_1\_5.0192463.pdf},
}

@article{Mittal2008,
author = {Mittal, R. and Dong, H. and Bozkurttas, M. and Najjar, F. M. and Vargas, A. and von Loebbecke, A.},
doi = {10.1016/j.jcp.2008.01.028},
issn = {10902716},
journal = {Journal of Computational Physics},
month = {may},
number = {10},
pages = {4825--4852},
pmid = {20216919},
publisher = {Academic Press Inc.},
title = {{A versatile sharp interface immersed boundary method for incompressible flows with complex boundaries}},
volume = {227},
year = {2008}
}

@article{WZhang2023,
author = {Zhang, Wei and Pan, Yu and Wang, Junshi and {Di Santo}, Valentina and Lauder, George V. and Dong, Haibo},
doi = {https://doi.org/10.1016/j.jcp.2023.111983},
issn = {0021-9991},
journal = {Journal of Computational Physics},
publisher = {Elsevier Inc.},
title = {{An Efficient Tree-Topological Local Mesh Refinement on Cartesian Grids for Multiple Moving Objects in Incompressible Flow}},
url = {https://doi.org/10.1016/j.jcp.2023.111983},
volume = {479},
year = {2023}
}

@article{Liu2017,
author = {Liu, Geng and Ren, Yan and Dong, Haibo and Akanyeti, Otar and Liao, James C. and Lauder, George V.},
doi = {10.1017/jfm.2017.533},
issn = {14697645},
journal = {Journal of Fluid Mechanics},
month = {oct},
pages = {65--88},
publisher = {Cambridge University Press},
title = {{Computational analysis of vortex dynamics and performance enhancement due to body-fin and fin-fin interactions in fish-like locomotion}},
volume = {829},
year = {2017}
}

@article{Buchholz2006,
author = {Buchholz, James H.J. and Smits, Alexander J.},
doi = {10.1017/S0022112005006865},
issn = {14697645},
journal = {Journal of Fluid Mechanics},
pages = {433--443},
title = {{On the evolution of the wake structure produced by a low-aspect-ratio pitching panel}},
volume = {546},
year = {2006}
}

@article{Menzer2025,
  title = {Fish schools in a vertical diamond formation: Effect of vertical spacing on hydrodynamic interactions},
  author = {Menzer, Alec and Pan, Yu and Lauder, George V. and Dong, Haibo},
  journal = {Phys. Rev. Fluids},
  volume = {10},
  issue = {4},
  pages = {043104},
  numpages = {29},
  year = {2025},
  month = {Apr},
  publisher = {American Physical Society},
  doi = {10.1103/PhysRevFluids.10.043104},
  url = {https://link.aps.org/doi/10.1103/PhysRevFluids.10.043104}
}

@article{ZhangLauder2024,
    doi = {10.1371/journal.pbio.3002501},
    author = {Zhang, Yangfan AND Ko, Hungtang AND Calicchia, Michael A. AND Ni, Rui AND Lauder, George V.},
    journal = {PLOS Biology},
    publisher = {Public Library of Science},
    title = {Collective movement of schooling fish reduces the costs of locomotion in turbulent conditions},
    year = {2024},
    month = {06},
    volume = {22},
    url = {https://doi.org/10.1371/journal.pbio.3002501},
    pages = {1-28},
    number = {6},

}

@article{YangfanPLOS2024,
    doi = {10.1371/journal.pbio.3002501},
    author = {Zhang, Yangfan AND Ko, Hungtang AND Calicchia, Michael A. AND Ni, Rui AND Lauder, George V.},
    journal = {PLOS Biology},
    publisher = {Public Library of Science},
    title = {Collective movement of schooling fish reduces the costs of locomotion in turbulent conditions},
    year = {2024},
    month = {06},
    volume = {22},
    url = {https://doi.org/10.1371/journal.pbio.3002501},
    pages = {1-28},
    number = {6},

}

@article{abrahams_risk_1985,
	title = {Risk of predation, hydrodynamic efficiency and their influence on school structure},
	volume = {13},
	issn = {1573-5133},
	url = {https://doi.org/10.1007/BF00000931},
	doi = {10.1007/BF00000931},
	language = {en},
	number = {3},
	journal = {Environmental Biology of Fishes},
	author = {Abrahams, Mark V. and Colgan, Patrick W.},
	month = jul,
	year = {1985},
	pages = {195--202}
}

@article{Hvas_2019,
	title = {Influence of experimental set-up and methodology for measurements of metabolic rates and critical swimming speed in {Atlantic} salmon {Salmo} salar},
	volume = {95},
	issn = {1095-8649},
	url = {https://pubmed.ncbi.nlm.nih.gov/31265133/},
	doi = {10.1111/jfb.14087},
	language = {en},
	number = {3},
	urldate = {2025-05-30},
	journal = {Journal of fish biology},
	author = {M, Hvas and F, Oppedal},
	month = sep,
	year = {2019},
	pmid = {31265133}
}

@article{partridge1982structure,
  title={The structure and function of fish schools},
  author={Partridge, Brian L},
  journal={Scientific american},
  volume={246},
  number={6},
  pages={114--123},
  year={1982},
  publisher={JSTOR}
}

@article{breder1967,
	title = {On the survival value of fish schools},
	journal = {Zoologica : scientific contributions of the New York Zoological Society},
	volume = {52},
	copyright = {In Copyright. Digitized with the permission of the rights holder},
	url = {https://www.biodiversitylibrary.org/part/203258},
	publisher = {New York, },
	author = {Breder, Charles M. (Charles Marcus)},
	year = {1967-09-22},
	pages = {25--40},
}

@article{handegard_method_2017,
	title = {Method to observe large scale behavioural waves propagating through fish schools using 4D sonar},
	volume = {74},
	issn = {1054-3139},
	url = {https://doi.org/10.1093/icesjms/fsw211},
	doi = {10.1093/icesjms/fsw211},
	number = {3},
	journal = {ICES Journal of Marine Science},
	author = {Handegard, Nils Olav and Holmin, Arne Johannes and Rieucau, Guillaume},
	month = mar,
	year = {2017},
	pages = {804--812}
}

@article{gerlotto_waves_2006,
	title = {Waves of agitation inside anchovy schools observed with multibeam sonar: a way to transmit information in response to predation},
	volume = {63},
	issn = {1054-3139},
	shorttitle = {Waves of agitation inside anchovy schools observed with multibeam sonar},
	url = {https://doi.org/10.1016/j.icesjms.2006.04.023},
	doi = {10.1016/j.icesjms.2006.04.023},
	number = {8},
	journal = {ICES Journal of Marine Science},
	author = {Gerlotto, François and Bertrand, Sophie and Bez, Nicolas and Gutierrez, Mariano},
	month = jan,
	year = {2006},
	pages = {1405--1417}
}

@article{makris_critical_2009,
	title = {Critical {Population} {Density} {Triggers} {Rapid} {Formation} of {Vast} {Oceanic} {Fish} {Shoals}},
	volume = {323},
	url = {https://www.science.org/doi/full/10.1126/science.1169441},
	doi = {10.1126/science.1169441},
	number = {5922},
	journal = {Science},
	author = {Makris, Nicholas C. and Ratilal, Purnima and Jagannathan, Srinivasan and Gong, Zheng and Andrews, Mark and Bertsatos, Ioannis and Godø, Olav Rune and Nero, Redwood W. and Jech, J. Michael},
	month = mar,
	year = {2009},
	pages = {1734--1737},
}

@article{Godin1985,
	author = {Godin, Jean-Guy J. and Morgan, M.Joanne},
	title = {Predator avoidance and school size in a cyprinodontid fish, the banded killifish (Fundulus diaphanus Lesueur)},
	year = {1985},
	journal = {Behavioral Ecology and Sociobiology},
	volume = {16},
	number = {2},
	pages = {105 – 110},
	doi = {10.1007/BF00295142},
	url = {https://www.scopus.com/inward/record.uri?eid=2-s2.0-0001214578&doi=10.1007%2fBF00295142&partnerID=40&md5=c023534057a8c176e55882b3b18310b7},
	publication_stage = {Final}
}

@Inbook{Pitcher1986,
author="Pitcher, Tony J.",
editor="Pitcher, Tony J.",
title="Functions of Shoaling Behaviour in Teleosts",
bookTitle="The Behaviour of Teleost Fishes",
year="1986",
publisher="Springer US",
address="Boston, MA",
pages="294--337",
isbn="978-1-4684-8261-4",
doi="10.1007/978-1-4684-8261-4_12",
url="https://doi.org/10.1007/978-1-4684-8261-4_12"
}

@article{Ranta1990,
 ISSN = {0003455X, 17972450},
 URL = {http://www.jstor.org/stable/23736020},
 author = {Esa Ranta and Kai Lindström},
 journal = {Annales Zoologici Fennici},
 number = {2},
 pages = {67--75},
 publisher = {Finnish Zoological and Botanical Publishing Board},
 title = {Assortative schooling in three-spined sticklebacks?},
 urldate = {2025-06-23},
 volume = {27},
 year = {1990}
}

@article{wolf_schooling_1987,
	title = {Schooling tendency and foraging benefit in the ocean surgeonfish},
	volume = {21},
	issn = {1432-0762},
	url = {https://doi.org/10.1007/BF00324436},
	doi = {10.1007/BF00324436},
	number = {1},
	journal = {Behavioral Ecology and Sociobiology},
	author = {Wolf, Nancy G.},
	month = jul,
	year = {1987},
	pages = {59--63},
}

@article{Shaw1978,
 ISSN = {00030996},
 URL = {http://www.jstor.org/stable/27848512},
 author = {Evelyn Shaw},
 journal = {American Scientist},
 number = {2},
 pages = {166--175},
 publisher = {Sigma Xi, The Scientific Research Society},
 title = {Schooling Fishes: The school, a truly egalitarian form of organization in which all members of the group are alike in influence, offers substantial benefits to its participants},
 urldate = {2025-06-23},
 volume = {66},
 year = {1978}
}

@article{kasumyan_schooling_2023,
	title = {Schooling {Behavior} of {Fish}: {General} {Ideas}, {Terms} and {Concepts}, {Prevalence}, {Applied} {Aspects}},
	volume = {63},
	issn = {1555-6425},
	url = {https://doi.org/10.1134/S0032945223070020},
	doi = {10.1134/S0032945223070020},
	number = {7},
	journal = {Journal of Ichthyology},
	author = {Kasumyan, A. O. and Pavlov, D. S.},
	month = dec,
	year = {2023},
	pages = {1219--1233},
}

@article{Neat2017,
author = {Neat, F. C.},
title = {Aggregating behaviour, social interactions and possible spawning in the deep-water fish Coryphaenoides rupestris},
journal = {Journal of Fish Biology},
volume = {91},
number = {3},
pages = {975-980},
doi = {https://doi.org/10.1111/jfb.13386},
url = {https://onlinelibrary.wiley.com/doi/abs/10.1111/jfb.13386},
eprint = {https://onlinelibrary.wiley.com/doi/pdf/10.1111/jfb.13386},
year = {2017}
}

@article{Lin_Wu_Zhang_Yang_2021, 
title={Flow-mediated organization of two freely flapping swimmers}, 
volume={912}, 
DOI={10.1017/jfm.2020.1143}, 
journal={Journal of Fluid Mechanics}, 
author={Lin, Xingjian and Wu, Jie and Zhang, Tongwei and Yang, Liming}, 
year={2021}, 
pages={A37}
}

@article{Lin_Wu_Zhang_Yang_2020, 
title={Self-organization of multiple self-propelling flapping foils: energy saving and increased speed}, 
volume={884}, 
DOI={10.1017/jfm.2019.954}, 
journal={Journal of Fluid Mechanics}, 
author={Lin, Xingjian and Wu, Jie and Zhang, Tongwei and Yang, Liming}, 
year={2020}, 
pages={R1}}

@article{Gungor_Khalid_Hemmati_2022, title={Classification of vortex patterns of oscillating foils in side-by-side configurations}, volume={951}, DOI={10.1017/jfm.2022.785}, journal={Journal of Fluid Mechanics}, author={Gungor, Ahmet and Khalid, Muhammad Saif Ullah and Hemmati, Arman}, year={2022}, pages={A37}}

@article{Han_2025, title={Tailoring formations of self-organising hydrofoil schools towards high-efficiency}, 
volume={1012}, 
DOI={10.1017/jfm.2025.10239}, 
journal={Journal of Fluid Mechanics}, 
author={Han, Tianjun and Mivehchi, Amin and Seyedmirzaei Sarraf, Seyedali and Moored, Keith W.}, 
year={2025}, 
pages={A26}}

@article{Guo_phalanx2025,
    author = {Guo, Jiacheng and Dong, Haibo},
    title = {A Computational Study of In-Phase and Anti-Phase Interactions of Fish in a Phalanx School},
    journal = {Journal of Fluids Engineering},
    volume = {147},
    number = {7},
    pages = {071106},
    year = {2025},
    month = {04},
    issn = {0098-2202},
    doi = {10.1115/1.4068037},
    url = {https://doi.org/10.1115/1.4068037},
    eprint = {https://asmedigitalcollection.asme.org/fluidsengineering/article-pdf/147/7/071106/7465580/fe\_147\_07\_071106.pdf},
}

\end{document}